\documentclass[12pt,preprint]{aastex}
\newcommand{\dif}[2]{\frac{\partial #1}{\partial #2}}

\usepackage{graphicx}
\usepackage{natbib}
\usepackage{amssymb}

\shorttitle{Formation of planetesimals}
\shortauthors{Michikoshi and Inutsuka}

\begin{document}
\title{The two-fluid analysis of the Kelvin-Helmholtz instability in dusty layer of a protoplanetary disk: A possible path toward the planetesimal formation through the gravitational instability}
\author{Shugo Michikoshi}
\affil{Department of Physics, Kyoto University, Kyoto 606-8502, Japan.}
\email{shugo@tap.scphys.kyoto-u.ac.jp}

\and

\author{Shu-ichiro Inutsuka}
\affil{Department of Physics, Kyoto University, Kyoto 606-8502, Japan.}
\email{inutsuka@tap.scphys.kyoto-u.ac.jp}

\begin{abstract}
We analyze the stability of dust layer in protoplanetary disk to understand the effect of the relative motion between gas and dust. The previous analyses not including the effect of relative motion between gas and dust show that the shear-induced turbulence may prevent the dust grains from settling sufficiently to be gravitationally unstable. We determine the growth rate of Kelvin-Helmholtz instability in wide range of parameter space, and propose a possible path toward the planetesimal formation through the gravitational instability. We expect the density of dust layer becomes $\rho_\mathrm{d}/\rho_\mathrm{g} \sim 100$ if the dust grains can grow up to 10m.
\end{abstract}

\section{Introduction}
The recent discoveries of extrasolar planets are indicating that the formation of Jupiter-mass planets is a common process.
According to the current standard model of planet formation, the growth of planet-sized body occurs in two sequential phases.
In the first phase, micron-sized dust grains grow up to the kilometer-sized bodies. The kilometer-sized bodies called planetesimals are the building blocks of terrestrial planets, cores of giant planets, comets, and asteroids. In the second stage, the planetesimals continue to grow through inelastic collisions.

The initial stage of coagulation of dust grains is supposed to continue up to about centimeter size.
However the growth of dust grains, from cm to km size, is not well understood. 
The agglomerative growth from submicron-sized to kilometer-sized bodies has 
many problems. The inward orbital drift associated with gas drag that would carry such meter-sized bodies from
1 AU into the central star is very rapid, and its timescale is only $10^2$ years \citep{Adachi1976,Weidenschilling1977}.
The direct assemblage of dust grains in laminar flow may be impossible.

One of the scenarios for this stage is due to gravitational instability (GI). If particles settle into a layer having sufficiently high density and low velocity dispersion, high density region spontaneously collapses under their self-gravity forming km-sized planetesimals \citep{Safronov1969,Goldreich1973,Coradini1981,Sekiya1983,Yamoto2004}.

There is a critical issue in this GI scenario. As dust grains settle toward the midplane, the rotational velocity around the midplane increases because of the reduced effect of the gas pressure compared to the centrifugal force and the solar gravity. The rotational velocity is a function of distance from the midplane, and the shear may induce Kelvin-Helmholtz instability (KHI). The slightest amount of turbulence in the nebular gas due to KHI prevents dust grains from settling.
Many authors have investigated this issue \citep{Weidenschilling1980,Cuzzi1993,Champney1995,Weidenschilling1995,Sekiya1998,Dobrovolskis1999}. In the case of the minimum mass solar nebula, they concluded that the gravitational fragmentation of the dust layer is impossible if the turbulence is developed. 
\citet{Sekiya1998} showed that large values of the total dust-to-gas mass ratio may provide a density cusp at the midplane.
\citet{Youdin2002} and \citet{Youdin2004} argued this cusp as a triggering of GI.

In order to understand the effect of instability in detail, linear stability analyses were performed.
\citet{Sekiya2000} have performed the linear analysis of the shear instability for the constant-Richardson-number dust density distribution. Their result confirms standard expectations by showing that the midplane shear layer becomes unstable when the minimum value of the Richardson number decreases below 0.226.  \citet{Sekiya2001} investigated the instability using the hybrid dust density distribution in which the dust layer has a constant density region and transition region.
For this distribution, the growth rate of the shear instability is much larger than the Kepler angular frequency when the dust density at the midplane is larger than the gas density.
\citet{Ishitsu2002} investigated the shear instability of the hybrid dust density distribution including the Coriolis force but neglecting the tidal force. Their results showed that the Coriolis force has little effect on the growth rate of the shear instability. \citet{Ishitsu2003} investigated the shear instability of the hybrid dust density distribution including both the Coriolis force and the tidal force. They showed that although the tidal force tends to stabilize the shear instability, the shear instability occurs before the dust density reaches the critical density of the gravitational instability.
Throughout those studies they use a single-fluid approximation. \citet{Garaud2004} performed the linear stability analysis based on two-fluid formalism with strong coupling approximation ($a<1\rm{cm}$) and destabilizing effect of radiative cooling.  They confirmed that the shear instability occurs prior to gravitational instability. 

In this paper, we remove the strong coupling approximation from the previous analyses, that is, we take into account the effect of the relative motion, neglecting the Coriolis force and the tidal force.
In Section \ref{sec:KH1} we investigate the effect of the relative motion between gas and dust in the two uniform fluids in relative motion separated by a horizontal boundary. We derive the condition of stabilization by the relative motion.
In Section \ref{sec:basiceq}, the basic equations for the linear analysis in a realistic protoplanetary disk are derived.
In Section \ref{sec:result}, results are given. In Section \ref{sec:discussion}, we discuss a possible path toward the planetesimal formation by the gravitational instability. Section \ref{sec:conclusion} is for conclusion.

\clearpage

\section{Kelvin-Helmholtz instabilities in the case of two uniform dusty fluids in relative horizontal motion \label{sec:KH1}}
The linear Kelvin-Helmholtz instability of shear layers is well-known for flows without dust grains \citep{Chandrasekhar1961}.
We consider the effect of dust in the Kelvin-Helmholtz instability.
First we examine the dispersion relation for the linear perturbation on Heaviside step function velocity profile in this section.
Since this velocity profile is very simple, one can obtain the analytic dispersion relation, which help
us to understand the character of this instability.
We analyze this dispersion relation to evaluate the stabilizing effect due to the inertia of dust. 
In the following sections we investigate the linear stability of the Kelvin-Helmholtz instability in a more realistic protoplanetary disk.

We suppose that the streaming takes place in the y-direction with the velocity $U(z)$, and 
gas and dust have the same velocity in the unperturbed state:
\begin{equation}
U(z)=
\left\{
	\begin{array}{rl}
		U_- & \mbox{for $z < 0 $} \\
		U_+ & \mbox{for $z > 0 $} 
	\end{array}
\right. .
\end{equation}
The unperturbed densities of the gas and the dust is uniform everywhere.
The gas is incompressible and inviscid fluid. The velocity dispersion of dust is negligible. For simplicity we do not take into account the influence of gravity in the vertical direction and the size distribution of dust grains.

The equations governing this system are
\begin{equation}
 \nabla \cdot \mathbf v_{\mathrm{g}} =0,
\end{equation}
\begin{equation}
\dif{\rho_{\rm{d}}}{t} + \nabla (\rho_{\rm{d}} \mathbf v_{\mathrm{d}}) =0,
\end{equation}
\begin{equation}
\rho_{\rm{g}} \left(\dif{\mathbf v_{\mathrm{g}}}{t}+(\mathbf v_{\mathrm{g}} \cdot \nabla)\mathbf v_{\mathrm{g}} \right)
=- \nabla P_g+ A \rho_{\rm{g}} \rho_{\rm{d}} (\mathbf v_{\mathrm{d}} - \mathbf v_{\mathrm{g}}) ,
\end{equation}
\begin{equation}
\rho_{\rm{d}} \left(\dif{\mathbf v_{\mathrm{d}}}{t}+(\mathbf v_{\mathrm{d}} \cdot \nabla)\mathbf v_{\mathrm{d}} \right)
= -A \rho_{\rm{g}} \rho_{\rm{d}} (\mathbf v_{\mathrm{d}} - \mathbf v_{\mathrm{g}}),
\end{equation}
where $A$ ,$P_g$,$\rho_{\rm{d}}$, $\rho_{\rm{g}}$, $\mathbf v_{\mathrm{d}}$, and $\mathbf v_{\mathrm{g}}$ are the friction coefficient, the gas pressure, the dust density, the gas density, the dust velocity, and the gas velocity, respectively.
We assume the form of perturbation quantities as
\begin{equation}
q(z) \exp (-i(\omega t-k y)).
\end{equation}
We only consider the perturbed motion in the $yz$-plane since these modes correspond to the most unstable modes of the Kelvin-Helmholtz instability.
To obtain the equations for the perturbation, we set the quantities 
$\mathbf U=(0,U(z),0)$,$\mathbf v_{\mathrm{g}1} = (v_{\rm{g}x1},v_{\rm{g}y1},v_{\rm{g}z1})$,$\mathbf v_{\mathrm{d}1} = (v_{\rm{d}x1},v_{\rm{d}y1},v_{\rm{d}z1})$,$\mathbf v_{\mathrm{g}}=\mathbf U+\mathbf v_{\mathrm{g}1}$,$\mathbf v_{\mathrm{d}}=\mathbf U+\mathbf v_{\mathrm{d}1}$, and
$P=P_0(z)+P_1(x,y,z)$.

The perturbation equations are
\begin{equation}
ikv_{\rm{g}y1}(z)+ \frac{dv_{\rm{g}z1}}{dz}(z)=0,
\label{eq:eomA1}
\end{equation}
\begin{equation}
- i (\omega- kU_{\pm} ) \rho_{\rm{d}1}(z) + i k \rho_{\rm{d}0} v_{\rm{d}y1}(z)  + \rho_{\rm{d}0} \frac{dv_{\rm{d}z1}}{dz}(z)=0,
\label{eq:eomA2}
\end{equation}
\begin{equation}
 -A \rho_{\rm{g}0} \rho_{\rm{d}0}(v_{\rm{d}x1}(z)- v_{\rm{g}x1}(z))-i\rho_{\rm{g}0} v_{\rm{g}x1}(z)( \omega - k U_{\pm})=0 ,
\label{eq:eomA3}
\end{equation}
\begin{equation}
 -A \rho_{\rm{g}0} \rho_{\rm{d}0}(v_{\rm{d}y1}(z)- v_{\rm{g}y1}(z))-i\rho_{\rm{g}0} v_{\rm{g}y1}(z)( \omega - k U_{\pm}) +i k P_1(z)=0,
\label{eq:eomA4}
\end{equation}
\begin{equation}
 -A \rho_{\rm{g}0} \rho_{\rm{d}0}(v_{\rm{d}z1}(z)-v_{\rm{g}z1}(z))-i\rho_{\rm{g}0} v_{\rm{g}z1}(z)( \omega - k U_{\pm}) + \frac{dP_1}{dz}(z)=0,
\label{eq:eomA5}
\end{equation}
\begin{equation}
 -A \rho_{\rm{d}0} \rho_{\rm{g}0}(v_{\rm{g}x1}(z)-v_{\rm{d}x1}(z))-i\rho_{\rm{d}0} v_{\rm{d}x1}(z)( \omega - k U_{\pm})=0 ,
\label{eq:eomA6}
\end{equation}
\begin{equation}
 -A \rho_{\rm{d}0} \rho_{\rm{g}0}(v_{\rm{g}y1}(z)-v_{\rm{d}y1}(z))-i\rho_{\rm{d}0} v_{\rm{d}y1}(z)( \omega - k U_{\pm})=0 ,
\label{eq:eomA7}
\end{equation}
\begin{equation}
 -A \rho_{\rm{d}0} \rho_{\rm{g}0}(v_{\rm{g}z1}(z)-v_{\rm{d}z1}(z))-i\rho_{\rm{d}0} v_{\rm{d}z1}(z)( \omega - k U_{\pm})=0 .
\label{eq:eomA8}
\end{equation}

We use the $x$-component of these equations (\ref{eq:eomA3}) and (\ref{eq:eomA6}), we have 
\begin{equation}
\omega - k U_{\pm} = - i A(\rho_{\rm{g}0}+\rho_{\rm{d}0}).
\end{equation}
This stable mode corresponds to a simple decay of the relative motion of the dust and the gas owing to the friction.

Eliminating the $v_{\rm{g}y1}(z),v_{\rm{d}y1}(z),v_{\rm{d}z1}(z),P_1(z)$ from the equations (\ref{eq:eomA1}),(\ref{eq:eomA4}),(\ref{eq:eomA5}),(\ref{eq:eomA7}), and (\ref{eq:eomA8})
, we have the equation of $v_{\rm{g}z1}(z)$:
\begin{equation}
\frac{d^2v_{\rm{g}z1}}{dz^2}(z) = k^2v_{\rm{g}z1}(z).
\label{eq:infeq}
\end{equation}
To obtain the condition that the perturbed quantities that do not diverge at $z=\pm \infty$, we have the solution of the equation (\ref{eq:infeq}):
\begin{equation}
v_{\rm{g}z1}(z)=
\left\{
	\begin{array}{rl}
		C_+ \exp (-kz) & \mbox{for $z > 0 $} \\
		C_- \exp (kz)  & \mbox{for $z < 0 $} 
	\end{array}
\right. ,
\end{equation}
where $C_+, C_-$ are arbitrary constants.

We need two boundary conditions. First we consider the boundary condition for velocity.
The z-component of the velocity vector $v_{\mathrm{g}z}$ at the interface is related to the position of the boundary $z_\mathrm b(t,x,y)$:
\begin{equation}
v_{\mathrm{g}z} = \frac{\partial z_\mathrm b}{\partial t}+ v_{\mathrm{g}x} \frac{\partial z_\mathrm b}{\partial x}+ v_{\mathrm{g}y} \frac{\partial z_\mathrm b}{\partial y}.
\end{equation}
The perturbation equation is given by
\begin{equation}
v_{\mathrm{g}z1} (x,y,\pm0,t)= U_\pm \frac{\partial z_\mathrm b}{\partial y} (x,y,t)+ \frac{\partial
z_\mathrm b}{\partial t} (x,y,t).
\end{equation}
We assume 
\begin{equation}
z_\mathrm b (x,y,t)= z_\mathrm b  \exp(-i(\omega t -k y)),
\end{equation}
and we have 
\begin{equation}
z_\mathrm b = \frac{v_{\mathrm{g}z1} (\pm0)}{i(k U_\pm - \omega)}.
\label{eq:eqofzb}
\end{equation}
We obtain the boundary condition that $v_{\mathrm{g}z1} (z)/(i(k U(z) - \omega))$ should be continuous at $z=0$. 
This condition leads to
\begin{equation}
\frac{C_+}{\omega - k U_{+}} = \frac{C_-}{\omega - k U_{-}}.
\end{equation}
However we do not need this boundary condition for dust (see Appendix A for the reason).

Second we must assume that the pressure is continuous at $(x,y,z_\mathrm{b}(x,y))$. $P_1(z)$ is given by
\begin{equation}
P_1(z)=i\frac{\rho_{\rm{g}0}}{k^2}(\omega-k U _\pm)  \frac{A(\rho_{\rm{g}0}+\rho_{\rm{d}0})-i(\omega-k U _\pm)}{A \rho_{\rm{g}0} - i(\omega-k U _\pm)} \frac{dv_{\rm{g}z1}}{dz}(z),
\label{eq:KH1:pre}
\end{equation}
and we obtain
\begin{equation}
-C_+(\omega-k U _+)  \frac{A(\rho_{\rm{g}0}+\rho_{\rm{d}0})-i(\omega-k U _+)}{A \rho_{\rm{g}0} - i(\omega-k U _+)}
=C_-(\omega-k U _-)  \frac{A(\rho_{\rm{g}0}+\rho_{\rm{d}0})-i(\omega-k U _-)}{A \rho_{\rm{g}0} - i(\omega-k U _-)}.
\end{equation}

With these conditions we can find the equation of $\omega$ by eliminating $C_{\pm}$ :
\begin{equation}
\omega^4
+i \frac{\mathcal{R}_\mathrm{d}+2}{t_{\rm{drag}}} \omega^3 - \frac{\mathcal{R}_\mathrm{d}+1}{t_{\rm{drag}}^2} \omega^2- \frac{i U^2 k^2 (\mathcal{R}_\mathrm{d}-2)}{4 t_{\rm{drag}}} \omega
- \frac{U^2 k^2 (\mathcal{R}_\mathrm{d}+1)}{4 t_{\rm{drag}}^2}-\frac{U^4 k^4}{16}=0,
\label{eq:eqforomega}
\end{equation}
where $\mathcal{R}_\mathrm{d}=\rho_{\rm{d}0}/\rho_{\rm{g}0}$, $t_{\rm{drag}}=1/A\rho_{\rm{g}0}$, and $U_\pm= \pm \frac{U}{2}$.

In the limit of either $\mathcal{R}_\mathrm{d} \to 0$, $t_{\rm{drag}} \to \infty$, or $t_{\rm{drag}} \to 0$, we have
\begin{equation}
\omega =\frac{k U}{2} i,
\end{equation}
which is the simple unstable mode of the original Kelvin-Helmholtz instability.
If there is no dust ($\mathcal{R}_\mathrm{d} \to 0$), obviously the dispersion relation reduces to that of the original Kelvin-Helmholtz instability.
If the friction is very small, the gas can move freely without the effect of dust, thus it reduces to the original Kelvin-Helmholtz instabilities. Furthermore, if the friction is very strong, the dust cannot move independently, that is, gas and dust behave as one strongly coupled fluid, and dispersion relation reduces again to the original Kelvin-Helmholtz instability.

The equation (\ref{eq:eqforomega}) depends on $k U$. We introduce $\omega= i \tilde \omega k U/2$ and $t_{\rm{drag}}=\tilde t_{\rm{drag}}/(k U)$, and we obtain the equation that does not depend on $k U$:
\begin{equation}
\tilde \omega^4
+ \frac{2(\mathcal{R}_\mathrm{d}+2)}{\tilde t_{\rm{drag}}} \tilde \omega^3+ \frac{4(\mathcal{R}_\mathrm{d}+1)}{{\tilde t_{\rm{drag}}}^2}\tilde \omega^2+ \frac{2 (\mathcal{R}_\mathrm{d}-2)}{\tilde t_{\rm{drag}}}\tilde \omega
- \frac{4(\mathcal{R}_\mathrm{d}+1)+{\tilde t_{\rm{drag}}}^2}{{\tilde t_{\rm{drag}}}^2}=0.
\label{eq:dr1}
\end{equation}
$\tilde \omega(\tilde t_{\rm{drag}},\mathcal{R}_\mathrm{d})$ determine the degree of instability.
In the limit of $t_{\rm{drag}}\to 0,\infty$, we have $\tilde \omega(\tilde t_{\rm{drag}},\mathcal{R}_\mathrm{d}) \to 1$.
This equation has always a real solution and we consider only a real solution (see Appendix \ref{apd:appr} for reason).

In Figure \ref{fig:contour_gasimcomp_step.eps}, we plot the contour of $\tilde \omega(\tilde t_{\rm{drag}},\mathcal{R}_\mathrm{d})$. The $\tilde \omega$ approaches to
 $1$ in $\mathcal{R}_{\mathrm{d}} \ll 1$ or $t_{\rm{drag}}\to \pm \infty$. It corresponds to the region where the friction is not effective for the reduction of growth rate. In the intermediate region, $\mathcal{R}_{\mathrm{d}} >1$ and $0 \ll t_{\rm{drag}} \ll \infty$, the growth rate is less than that of the original Kelvin-Helmholtz Instability.

We consider the condition of stabilization. The friction should not be strong or weak to stabilize Kelvin-Helmholtz instabilities. We express this condition as $\rm{min}(1/A \rho_{\rm{g}},1/A \rho_{\rm{d}}) <1/k U < \rm{max}(1/A \rho_{\rm{g}},1/A \rho_{\rm{d}}) $.
We need also that the amount of dust is more than that of gas, $\mathcal{R}_\mathrm{d}>1$.

The growth rate is reduced less than $\mu$ ($\tilde \omega < \mu$) if the condition is satisfied, which is
\begin{equation}
C_{\mathrm{reduce}1}  < k U t_{\mathrm{drag}} < C_{\mathrm{reduce}2}
\label{eq:condstab}
\end{equation}
where 
\begin{equation}
C_{\mathrm{reduce}1} =\frac{\left( \mathcal{R}_\mathrm{d} -2 \right) \mu  + \left( \mathcal{R}_\mathrm{d}+ 2 \right) {\mu }^3 -
      {\sqrt{{\mathcal{R}_\mathrm{d} }^2 \mu^2 {\left( 1  + \mu^2 \right) }^2 - 
          4\mathcal{R}_\mathrm{d} \left( 1 - {\mu }^4 \right) -4{\left( 1 - {\mu }^2 \right) }^2 }
          }}{1- {\mu }^4},
\end{equation}
\begin{equation}
C_{\mathrm{reduce}2} =\frac{\left( \mathcal{R}_\mathrm{d} -2 \right) \mu  + \left( \mathcal{R}_\mathrm{d}+ 2 \right) {\mu }^3 + 
      {\sqrt{{\mathcal{R}_\mathrm{d} }^2 \mu^2 {\left( 1  + \mu^2 \right) }^2 - 
          4\mathcal{R}_\mathrm{d} \left( 1 - {\mu }^4 \right) -4{\left( 1 - {\mu }^2 \right) }^2 }
          }}{1- {\mu }^4} .
\end{equation}
In the case of $ \mathcal{R}_{\mathrm{d}} \gg 1$, we have the approximated condition:
\begin{equation}
\frac{2(1-\mu^2)}{\mu(1+\mu^2)} < k U t_{\mathrm{drag}} < \frac{2\mu}{1-\mu^2} \mathcal{R}_{\mathrm{d}}.
\end{equation}
In addition, in the case of $\mu \ll 1$, we have the simple condition:
\begin{equation}
\frac{2}{\mu} < k U t_{\mathrm{drag}} < 2\mu  \mathcal{R}_{\mathrm{d}}.
\end{equation}
Figure \ref{fig:contour_gasimcomp_step.eps} shows that the growth rate in the region satisfying this condition is reduced.

\begin{figure}
\plotone{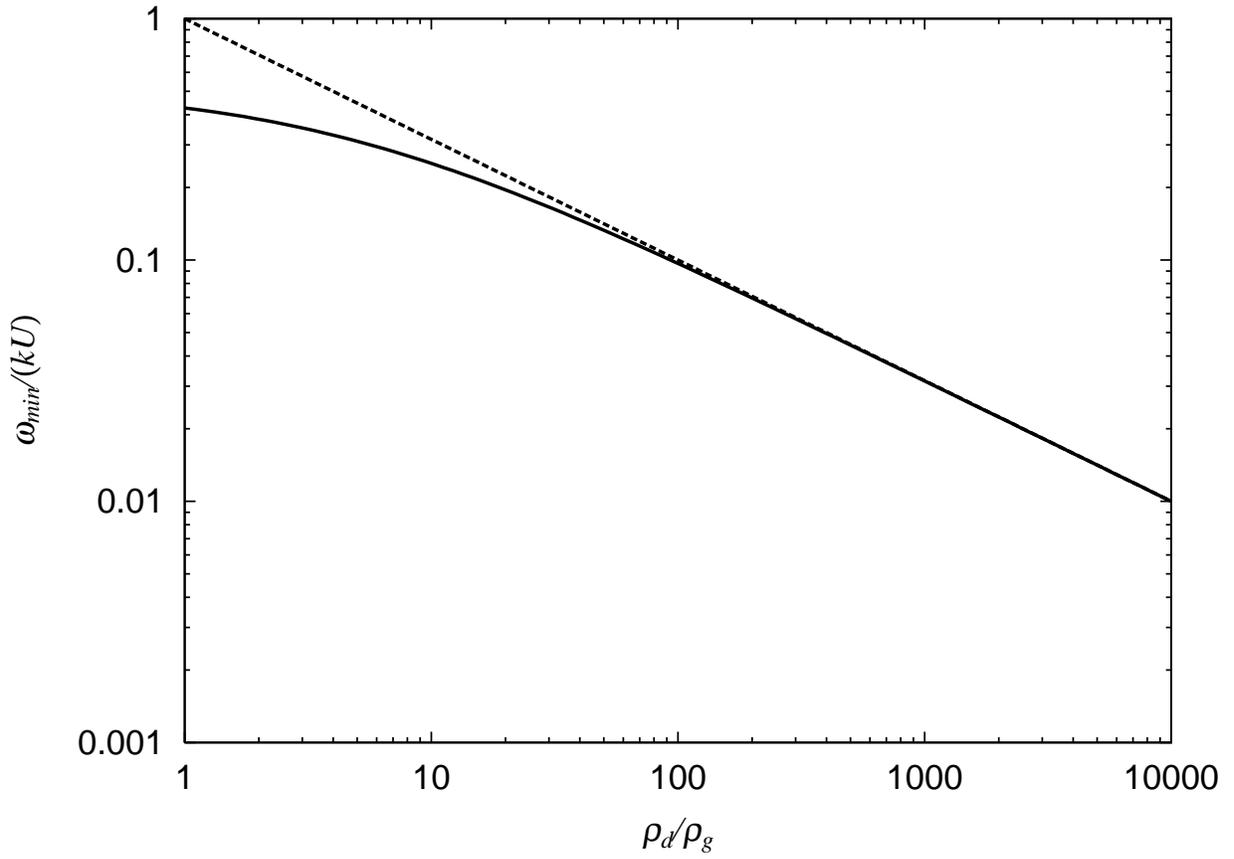}
\caption{The maximum stabilized growth rate. The solid lines denotes the numerical solution, and the dashed lined denotes approximate solution.}
\label{fig:shearminimum.eps}
\end{figure}

We study how the friction reduces the growth rate of the Kelvin-Helmholtz instabilities.
We seek the minimum of $\tilde \omega(\tilde t_{\rm{drag}},\mathcal{R}_\mathrm{d})$ in $\mathcal{R}_\mathrm{d}=\rm{const.}$. In the case of $\mathcal{R}_\mathrm{d} \gg 1$, one can obtain 
\begin{equation}
\omega_{\rm{min}}(\mathcal{R}_{\mathrm{d}}) \simeq \frac{2}{\sqrt{\mathcal{R}_{\mathrm{d}} } }\frac{k U}{2} i,
\end{equation}
\begin{equation}
t_{\rm{drag}} \simeq 2\sqrt{\mathcal{R}_{\mathrm{d}} } \frac{1}{k U}.
\end{equation}
As shown in Figure \ref{fig:shearminimum.eps}, this expression is in good agreement with the numerical solution for $\mathcal{R}_\mathrm{d}>10$.

\begin{figure}
\plotone{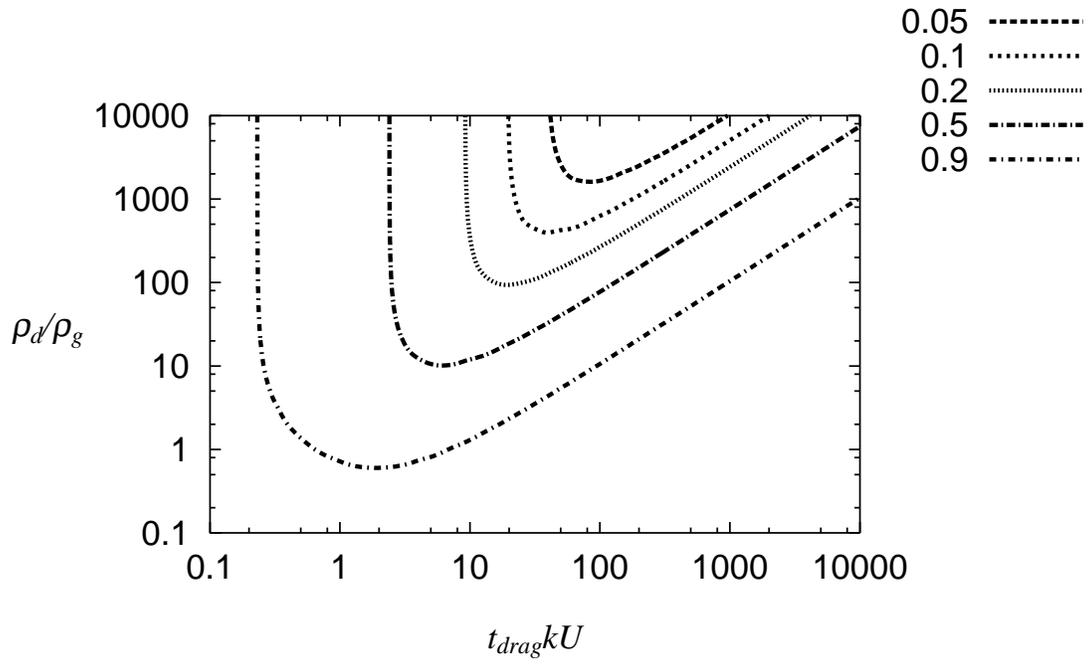}
\caption{The contour of growth rate $\tilde \omega$ as a function of $\rho_\mathrm{d}/\rho_\mathrm{g}$ and $t_{\mathrm{drag}}$.}
\label{fig:contour_gasimcomp_step.eps}
\end{figure}

\clearpage
\section{The case of dusty layer between two gaseous layers\label{sec:KH2}}
In this section we consider the effect of the thickness of the dust layer. 
The dusty gas layer whose thickness $2d$ is sandwiched between two gas layers without dust.
For simplicity we assume the unperturbed state is symmetric with respect to the equatorial plane ($U(z)=U(-z), \rho_{\mathrm{g}0}(z)=\rho_{\mathrm{g}0}(-z), \rho_{\mathrm{d}0}(z)=\rho_{\mathrm{d}0}(-z)$):
\begin{equation}
U(z)=
\left\{
	\begin{array}{rl}
		U & \mbox{for $|z| < d $} \\
		0 & \mbox{for $|z| > d $} 
	\end{array}
\right. ,
\end{equation}
\begin{equation}
\rho_{\mathrm{d}0}(z)=
\left\{
	\begin{array}{rl}
	\rho_{\rm{d}0} & \mbox{for $|z| < d $} \\
		0 & \mbox{for $|z| > d $} 
	\end{array}
\right. ,
\end{equation}
\begin{equation}
\rho_{\rm{g}0}(z)=\rho_{\rm{g}0}.
\end{equation}

The perturbation equations are the same in Section \ref{sec:KH1}.
Thus we have the perturbed velocity that does not diverge at $z=\infty$:
\begin{equation}
v_{\rm{g}z1}(z)=
\left\{
	\begin{array}{rl}
	C_1 e^{kz} + C_2 e^{-kz} & \mbox{for $0< z < d $} \\
		C_3 e^{-kz} & \mbox{for $d< z$} 
	\end{array}
\right. ,
\end{equation}
where $C_1, C_2,$ and $C_3$ are constant.

From (\ref{eq:eqofzb}) and (\ref{eq:KH1:pre}), we have two boundary conditions at $z=z_\mathrm{d}$:
\begin{eqnarray}
(e^{2kd} C_1  - C_2) \frac{(k U - \omega)(kU -\omega - i A (\rho_{\mathrm{d}0}+\rho_{\mathrm{g}0}))}{k U -\omega - i A\rho_{\mathrm{g}0}}-   \omega C_3&=&0 
\label{eq1:KH2_1}
\\
e^{2kd} \omega C_1 +\omega C_2 -( \omega - k U) C_3&=&0.\label{eq1:KH2_2}
\end{eqnarray}
Because of the symmetry of the unperturbed state, there are two types of solutions: even solutions where $P_1(z)=P_1(-z)$, $v_{\rm{g}z1}(z)=-v_{\rm{g}z1}(-z)$ and odd solutions where $P_1(z)=-P_1(-z)$, $v_{\rm{g}z1}(z)=v_{\rm{g}z1}(-z)$.
The boundary conditions at $z=0$ are written as
\begin{equation}
C_1+S C_2=0,\label{eq1:KH2_3}
\end{equation}
where $S=\pm1$.
The solutions in the case of $S=1$ correspond to the even solutions, the solutions in the case of $S=-1$ correspond to the odd solutions

By the condition that equation (\ref{eq1:KH2_1}), equation (\ref{eq1:KH2_2}), and equation (\ref{eq1:KH2_3}) have a nontrivial solution, we obtain the dispersion relation:
\begin{eqnarray}
&&\mu^3+
  \left( \mathcal{R}_\mathrm{d}e^{-2kd} {\tilde t_{\mathrm{drag}}}^{-1}S  + \left( 2 + \mathcal{R}_\mathrm{d}  \right) {\tilde t_{\mathrm{drag}}^{-1}}  + 
       2i\left( S e^{-2kd} + 2 \right) \right)\mu^2 \nonumber \\
&+& 2i  \left( e^{-2kd}S + 1 \right) \left( 2(1 + \mathcal{R}_\mathrm{d}){\tilde t_{\mathrm{drag}}}^{-1}  + 3i  \right)   \mu 
 -4\left( e^{-2kd}S + 1\right) \left( (1 + \mathcal{R}_\mathrm{d}){\tilde t_{\mathrm{drag}}}^{-1}  + i  \right)  =0,
\end{eqnarray}
where $\mu =  \omega/(k Ui/2)$, $\tilde t_{\mathrm{drag}}=t_{\mathrm{drag}}k U $, $\mathcal{R}_\mathrm{d}=\rho_{\mathrm{d}0}/\rho_{\mathrm{g}0}$.

If the friction is very weak ( $\tilde t_{\mathrm{drag}} \gg 1$ ) , the growth rate $\mathrm{Im} (\omega)$ is given by
\begin{equation}
\mathrm{Im} (\omega)= \frac{k U}{2} \sqrt{1-e^{-4kd}}.
\end{equation}
In this case, the odd mode and even mode have the same growth rate.

If the friction is very strong, the growth rate $\mathrm{Im} (\omega)$ is given by
\begin{equation}
\mathrm{Im}(\omega)=
\left\{
	\begin{array}{rl}
	\displaystyle{kU \frac{\sqrt{(1+\mathcal{R}_\mathrm{d})(1-e^{-4k d})}}{\mathcal{R}_\mathrm{d}(1+ e^{-2k d})+2}} & \mbox{(even mode)} \\
	\displaystyle{kU \frac{\sqrt{(1+\mathcal{R}_\mathrm{d})(1-e^{-4k d})}}{\mathcal{R}_\mathrm{d}(1- e^{-2k d})+2}} & \mbox{(odd mode)} 
	\end{array}
\right. .
\label{eq:KH2:SC}
\end{equation}
Thus odd mode is more unstable than even mode. In the limit of $\mathcal{R}_\mathrm{d} \to 0$, 
the growth rate for these modes approaches to the growth rate for the case of weak friction.

In the long wavelength limit ($k d\to 0$), 
one can show 
\begin{equation}
\mu =
\left\{
	\begin{array}{rl}
\displaystyle{
	 \frac{2}{\sqrt{1+\mathcal{R}_\mathrm{d}}} (kd)^{1/2}- 2i}  & \mbox{(even mode)} \\
	c_1 (kd)^{1/2} - i c_2 (kd)^{1/2}  & \mbox{(odd mode)} 
	\end{array}
\right. ,
\label{eq:KH2:LW}
\end{equation}
where $c_1$ and  $c_2$ are given by
\begin{equation}
c_1^2-c_2^2=\frac{4(1+\mathcal{R}_\mathrm{d}+\tilde t_{\mathrm{drag}}^2)}{1+\tilde t_{\mathrm{drag}}^2}
\label{eq:KH2a}
\end{equation}
\begin{equation}
c_1c_2=\frac{2 \mathcal{R}_\mathrm{d} \tilde t_{\mathrm{drag}}}{1+\tilde t_{\mathrm{drag}}^2}.
\label{eq:KH2b}
\end{equation}
From the equation (\ref{eq:KH2a}), we give the inequality:
\begin{equation}
c_1^2 > 4 \frac{\mathcal{R}_\mathrm{d} +1+\tilde t_{\mathrm{drag}}^2}{1+\tilde t_{\mathrm{drag}}^2}=4 \left(1+ \frac{\mathcal{R}_\mathrm{d}}{1+\tilde t_{\mathrm{drag}}^2} \right) > 4 > \frac{4}{1+\mathcal{R}_\mathrm{d}}.
\end{equation}
This inequality means that the odd mode is always more unstable than the even mode in the case of the long wavelength limit.
We can obtain the analytic solutions of $c_1$ and $c_2$:
\begin{eqnarray}
c_1^2&=&2 \sqrt{\frac{(\mathcal{R}_\mathrm{d}+1)^2+\tilde t_{\mathrm{drag}}^2}{1+\tilde t_{\mathrm{drag}}^2}}+2\frac{\mathcal{R}_\mathrm{d}+1+\tilde t_{\mathrm{drag}}^2}{1+\tilde t_{\mathrm{drag}}^2}
\end{eqnarray}
\begin{eqnarray}
c_2^2&=&2 \sqrt{\frac{(\mathcal{R}_\mathrm{d}+1)^2+\tilde t_{\mathrm{drag}}^2}{1+\tilde t_{\mathrm{drag}}^2}}-2\frac{\mathcal{R}_\mathrm{d}+1+\tilde t_{\mathrm{drag}}^2}{1+\tilde t_{\mathrm{drag}}^2}
\end{eqnarray}

In the short wavelength limit ($k d\to \infty$), $\mu$ is constant for $k$. There is no difference of the growth rate between the even mode and the odd mode.

The growth rate is proportional to
$k$ ($k d\to \infty$) or $k^{3/2}$ ($k d\to 0$). In the short wavelength limit the growth rate diverges owing to the discontinuity of the unperturbed velocity. In the Section \ref{sec:result},we show that the growth rate does not diverge in the short wavelength limit if the unperturbed velocity is continuous.

\begin{figure}
\plotone{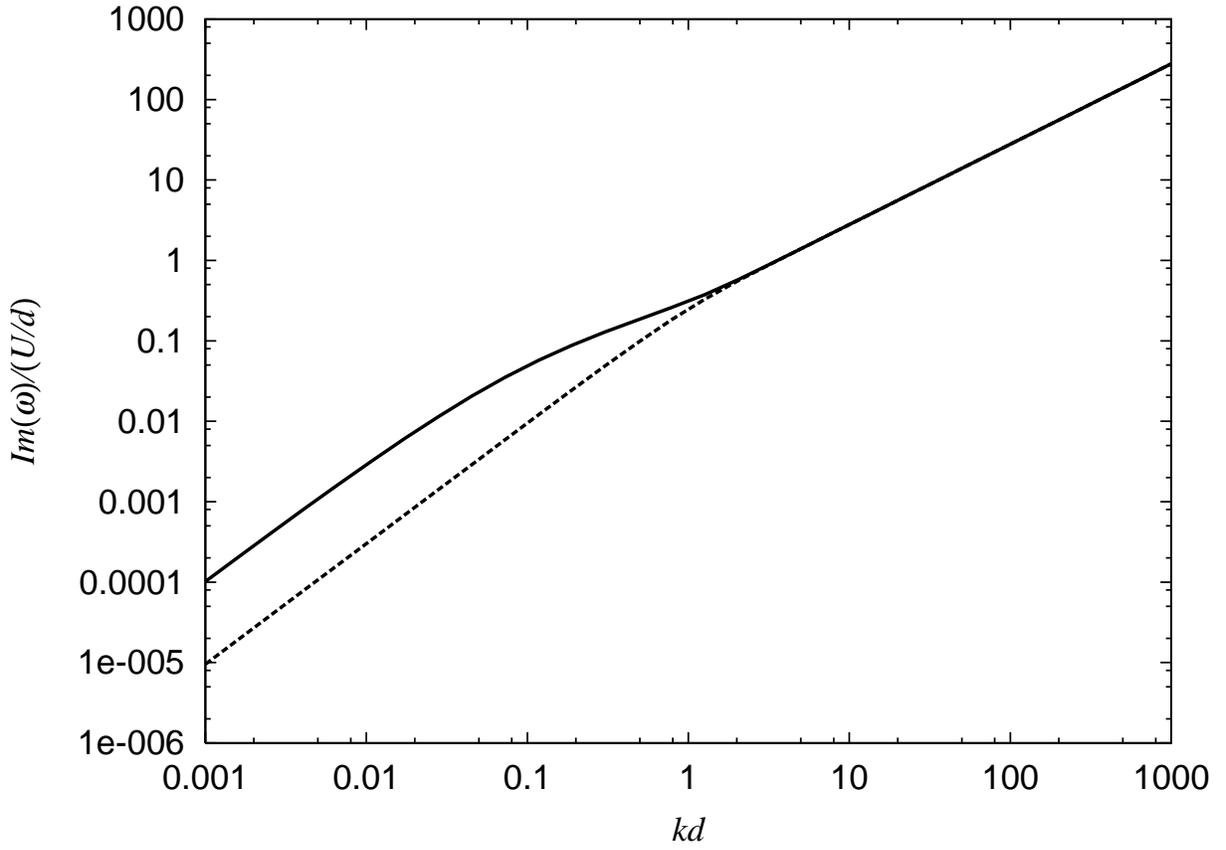}
\caption{The growth rate as functions of wavenumber $k$ for odd mode (solid curve) and even mode (dotted curve) in the case of $t_{\mathrm{drag}} =0.1 U/d$, $\mathcal{R}_\mathrm{d}=10$.
}
\label{fig:three_layer.eps}
\end{figure}

In Figure \ref{fig:three_layer.eps} we show the dispersion relation in the case of $t_{\mathrm{drag}} =0.1 U/d$, $\mathcal{R}_\mathrm{d}=10$. In the case of $k d<1$, there is the difference between the even mode and the odd mode, and the growth rate is proportional to $k^{3/2}$. In the case of $k d>1$, the even mode and the odd mode have similar growth rate, and the growth rate is proportional to $k$.

\clearpage

\section{Basic equations\label{sec:basiceq}}
In this section we analyze the stability of a structured dust layer in the protoplanetary disk, which is more realistic than the step function velocity profile studied in the previous section.

We do not assume that the dust aggregates are small enough to couple firmly: the dust aggregates can move relatively to the gas.
For simplicity, we use the local Cartesian coordinates $(x,y,z)$ and neglect the curvature of the orbit of the perturbed state, and do not consider the gravity of central star and the effect of Colioris force.
The equations governing this system are
\begin{equation}
 \nabla \cdot \mathbf v_{\mathrm{g}} =0,
\end{equation}
\begin{equation}
\dif{\rho_{\rm{d}}}{t} + \nabla (\rho_{\rm{d}} \mathbf v_{\mathrm{d}}) =0,
\end{equation}
\begin{equation}
\rho_{\rm{g}} \left(\dif{\mathbf v_{\mathrm{g}}}{t}+(\mathbf v_{\mathrm{g}} \cdot \nabla)\mathbf v_{\mathrm{g}} \right)
=- \nabla P_g+ A \rho_{\rm{g}} \rho_{\rm{d}} (\mathbf v_{\mathrm{d}} - \mathbf v_{\mathrm{g}}) ,
\end{equation}
\begin{equation}
\rho_{\rm{d}} \left(\dif{\mathbf v_{\mathrm{d}}}{t}+(\mathbf v_{\mathrm{d}} \cdot \nabla)\mathbf v_{\mathrm{d}} \right)
= -A \rho_{\rm{g}} \rho_{\rm{d}} (\mathbf v_{\mathrm{d}} - \mathbf v_{\mathrm{g}}).
\end{equation}

We assume the gas density $\rho_{\rm{g0}}$ is uniform ($d\rho_{\rm{g}0}/dz=0$) in the dust layer since we suppose a thin dust layer.
We adopt the model of dust density $\rho_{\rm{d}0}(z)$ which has the form as
\begin{equation}
\rho_{\rm{d}0}(z) =
\left\{
	\begin{array}{cr}
		 \rho_{\rm{d}0} & \mbox{for $|z| <z_{\rm{d}}-2h_{\rm{d}} $} \\
	\displaystyle{
		 \frac{\rho_{\rm{d}0}}{2} \left[1- \sin \pi \frac{z- z_{\rm{d}}+h_{\rm{d}}}{2 h_{\rm{d}}} \right]} & \mbox{for $z_{\rm{d}}-2h_{\rm{d}} <|z| < z_{\rm{d}} $} \\
		 0 & \mbox{for $z_{\rm{d}} < z $} 
	\end{array}
\right. ,
\end{equation}
where $z_{\rm{d}}$ the half thickness of the dust layer,
$h_{\rm{d}}$ the half thickness of the transition layer, $z$ the distance from the midplane, and $\rho_{\rm{d}0}$ the dust density at the midplane \citep{Sekiya2001}. $z_{\rm{d}}$ is given by
\begin{equation}
z_{\rm{d}} = \frac{\Sigma_{\rm{d}}}{2\rho_{\rm{d}0}}+h_{\rm{d}},
\end{equation}
where $\Sigma_{\rm{d}}=\int_{-\infty}^{\infty} \rho_{\mathrm{d}}(z) dz$ is the surface density of dust.

We use here a frame of reference moving with gas dominant region ($\rho_{\rm{d}0}(z)=0$).
The stationary rotational velocity of gas is given by \citet{Nakagawa1986}:
\begin{equation}
U(z) = \frac{\mathcal{R}_\mathrm{d}(z) (1+\mathcal{R}_\mathrm{d}(z)) \Gamma^2}{1+(1+\mathcal{R}_\mathrm{d}(z))^2 \Gamma^2} \eta v_{\rm{k}},
\label{eq:steady_velo}
\end{equation}
where $\mathcal{R}_\mathrm{d}(z)=\rho_{\rm{d}0}(z) / \rho_{\rm{g}0}$, $\Gamma=\rho_{\rm{g}} A/\Omega_{\rm{k}}$, $v_{\rm{k}}$  the circular Kepler velocity, $\Omega_{\rm{k}}$ Kepler frequency, $A$ the friction coefficient.
$\eta$ is a non-dimensional parameter which represents the effect of the radial pressure gradient:
\begin{equation}
\eta = - \frac{c_{\rm{s}}^2}{2v_{\rm{k}}^2} \dif{\log P}{\log r}
\label{eq:eta},
\end{equation}
where $P$ is the gas pressure, $c_{\rm{s}}$ the isothermal sound velocity, and $r$ the distance from the rotational axis.
We assume that the friction coefficients for particles whose sizes are smaller than the mean free path of the gas are determined by the Epstein's drag law, and those larger than mean free path are determined by the Stokes law: 
\begin{equation}
A=
\left\{
	\begin{array}{rl}
	\displaystyle{
		\frac{c_{\rm{t}}}{\rho_{\rm{mat}} a}} & \mbox{for $\displaystyle{a < \frac{3}{2}\lambda_{\rm{mfp}}} $} \\
	\displaystyle{
		\frac{3 \lambda_{\rm{mfp}} c_{\rm{t}}}{2 \rho_{\rm{mat}} a^2}} & \mbox{for $\displaystyle{a > \frac{3}{2} \lambda_{\rm{mfp}}} $} 
	\end{array}
\right. ,
\end{equation}
where $\lambda_{\rm{mfp}}$ is mean free path of the gas, $a$ the radius of dust aggregate, $\rho_{\rm{mat}}$ the internal mass density of dust aggregates, and $c_{\rm{t}}$ the thermal velocity of gas. We set $\rho_{\rm{mat}}=1\rm{g} \rm{cm}^{-3}$.
For the sake of simplicity, we neglect the radial velocity of gas, and assume that the dust and the gas have the same velocity in the unperturbed state.
The equations are similar to those of Section \ref{sec:KH1}, except that the unperturbed velocity and the unperturbed dust density is not uniform.
We restrict ourselves to the case where $k_x=0$ since this gives most unstable mode.
We have the perturbed equation:
\begin{equation}
\frac{d^2v_{\rm{g}z1}}{dz^2}=C_1(U(z),\rho_{\rm{d}0}(z),\rho_{\rm{g}},\omega,k,A) \frac{dv_{\rm{g}z1}}{dz}
+C_2(U(z),\rho_{\rm{d}0}(z),\rho_{\rm{g}},\omega,k,A)v_{\rm{g}z1}.
\label{eq:zeq}
\end{equation}
 $C_1$ and $C_2$ are
\begin{eqnarray}
C_1 &=& -\frac{\frac{d \mathcal{R}_\mathrm{d}}{dz}(z)}{1+\mathcal{R}_\mathrm{d} - i t_{\rm{drag}} \omega_0} ,\\
C_2 &=& -\frac{k \left( i  + t_{\rm{drag}} \omega _0 \right)
}{{{\omega }_0} 
    \left( 1 + \mathcal{R}_\mathrm{d}  - i  {t_{\rm{drag}}} {{\omega }_0} \right) } \times \nonumber \\
&&
 i\left(    k {{\omega }_0} - U''(z)  +
 \frac{  \mathcal{R}_\mathrm{d}'  U'(z)+ \mathcal{R}_\mathrm{d}   U''(z)}{{\left( i  + {t_{\rm{drag}}} {{\omega }_0} \right) }^2} +
 \frac{  i \mathcal{R}_\mathrm{d} k {{\omega }_0}} {{\left( i  + {t_{\rm{drag}}} {{\omega }_0} \right) }}   + 
           \frac{ 2  k\mathcal{R}_\mathrm{d} {t_{\rm{drag}}} {U'(z)}^2   }
         {{\left( i  + {t_{\rm{drag}}} {{\omega }_0} \right) }^3} \right) ,
\end{eqnarray}
where a prime means the derivative with respect to $z$,
\begin{equation}
\omega_0=\omega-k U(z),
\end{equation}
\begin{equation}
t_{\rm{drag}}=\frac{1}{A\rho_{\rm{g}0}},
\end{equation}
In the limit of $t_{\rm{drag}} \to 0$ this equation approaches to equation (22) of \citet{Sekiya2000} without the effect of gravity.
The perturbed quantities are given by
\begin{eqnarray}
P_1&=&\frac{\rho_{\rm{g}0} \omega_0 (-1 - \mathcal{R}_\mathrm{d} + i t_{\rm{drag}} \omega_0)}{k^2(i+t_{\rm{drag}} \omega_0)} \frac{dv_{\rm{g}z1}}{dz}
+\frac{i U'(z) \rho_{\rm{g}0} \left(1-\frac{\mathcal{R}_\mathrm{d}}{(i+t_{\rm{drag}} \omega_0)^2}\right)}{k}v_{\rm{g}z1}
\label{eq:press},\\
v_{\rm{g}y1}&=& \frac{i}{k}\frac{dv_{\rm{g}z1}}{dz}(z),\\
v_{\rm{d}y1}&=& \frac{- i {v_{\rm{g}z1}}'(z)+t_{\rm{drag}}(k v_{\rm{g}z1}(z) U'(z)-\omega_0 {v_{\rm{g}z1}}'(z))}{k(i+t_{\rm{drag}} \omega_0)^2},
\end{eqnarray}
\begin{equation}
v_{\rm{d}z1}= \frac{i v_{\rm{g}z1}(z)}{i+t_{\rm{drag}} \omega_0}.
\end{equation}

In region outside the dust layer $z>z_{\rm{d}}$, equation (\ref{eq:zeq}) becomes
\begin{equation}
\frac{d^2v_{\rm{g}z1}}{dz^2}=k^2 v_{\rm{g}z1}.
\end{equation}
We find the solution of the perturbed quantity that does not diverge at $z\to \infty$:
\begin{equation}
v_{\rm{g}\it{z}1}(z)=
	\begin{array}{rl}
		C_+ \exp (-kz) & \mbox{for $z > z_{\rm{d}} $} 
	\end{array},
\label{eq:extsol}
\end{equation}
where $C_+$ is a constant. 

In the case of $z<z_{\rm{d}}$, we cannot obtain the analytic solution.
Thus we solve equation (\ref{eq:zeq}) numerically.

We consider the boundary conditions.
First $v_{\mathrm{g}z1}(z)/(\omega - k U(z))$ should be continuous at $z=z_{\rm{d}}$:
\begin{equation}
\frac{v_{\rm{g}z1}(z_{\rm{d}}+0)}{\omega-k U(z_{\rm{d}}+0)} = \frac{v_{\rm{g}z1}(z_{\rm{d}}-0)}{\omega-k U(z_{\rm{d}}-0)}.
\end{equation}
Since the unperturbed velocity is continuous, this condition means that $v_{\rm{g}z1}$ is continuous at $z=z_{\rm{d}}$:
\begin{equation}
v_{\rm{g}z1} (z_{\rm{d}}+0)=v_{\rm{g}z1} (z_{\rm{d}}-0)
\label{eq:extbound1}
\end{equation}
Second the gas pressure should be continuous. With equation (\ref{eq:press}) , (\ref{eq:extsol}), and (\ref{eq:extbound1}), we obtain the boundary condition at $z=z_{\rm{d}}$:
\begin{equation}
k v_{\rm{g}z1}(z_{\rm{d}}-0) + \frac{d v_{\rm{g}z1}}{dz}(z_{\rm{d}}-0)=0.
\label{eq:extboundA}
\end{equation}

There are two types of solutions: even solutions where $P_1(z)=P_1(-z)$, $v_{\rm{g}z1}(z)=-v_{\rm{g}z1}(-z)$ and odd solutions where $P_1(z)=-P_1(-z)$, $v_{\rm{g}z1}(z)=v_{\rm{g}z1}(-z)$.
We have the boundary condition at z=0:
\begin{equation}
\left\{
	\begin{array}{c r}
		 v_{\mathrm{g}\mathit{z}1} = 0 & \mbox{(even mode)} \\
		 \displaystyle{\frac{d v_{\mathrm{g}\mathit{z}1}}{dz}} = 0 & \mbox{(odd mode)} 
	\end{array}
\right. ,
\label{eq:intboundB}
\end{equation}

In the next section we solve equation (\ref{eq:zeq}) with the boundary condition (\ref{eq:extboundA}) and (\ref{eq:intboundB}) numerically.

\clearpage

\section{Numerical Result\label{sec:result}}
We adopt the model of the minimum solar nebula at $r=1\rm{AU}$ \citep{Hayashi1981}:
\begin{equation}
\Sigma_{d}=
\left\{
	\begin{array}{rl}
		7.1 \left(\frac{r}{\rm{AU}} \right)^{-3/2} \rm{g}/\rm{cm}^2 & \mbox{for ($0.3 \rm{AU} <r < 2.7\rm{AU}$) } \\
		30 \left(\frac{r}{\rm{AU}} \right)^{-3/2}\rm{g}/\rm{cm}^2  & \mbox{for $ (2.7 \rm{AU} < r < 36\rm{AU})$} 
	\end{array}
\right. ,
\end{equation}
\begin{eqnarray}
\eta = 1.81 \times 10^{-3} \left( \frac{r}{\rm{AU}}\right)^{1/2},
\end{eqnarray}
\begin{eqnarray}
\rho_{\rm{g}0} = 1.4 \times 10^{-9} \left(\frac{r}{1\rm{AU}} \right)^{-11/4},
\end{eqnarray}
\begin{equation}
\lambda_{\rm{mfp}}=1.4 \left(\frac{r}{\rm{AU}} \right)^{11/4} \rm{cm},
\end{equation}
\begin{equation}
c_{\rm{s}} = 1.1 \times 10^5 \left( \frac{r}{1\rm{AU}} \right) ^{-1/4} \rm{cm} \cdot s^{-1}.
\end{equation}

\begin{figure}
\plottwo{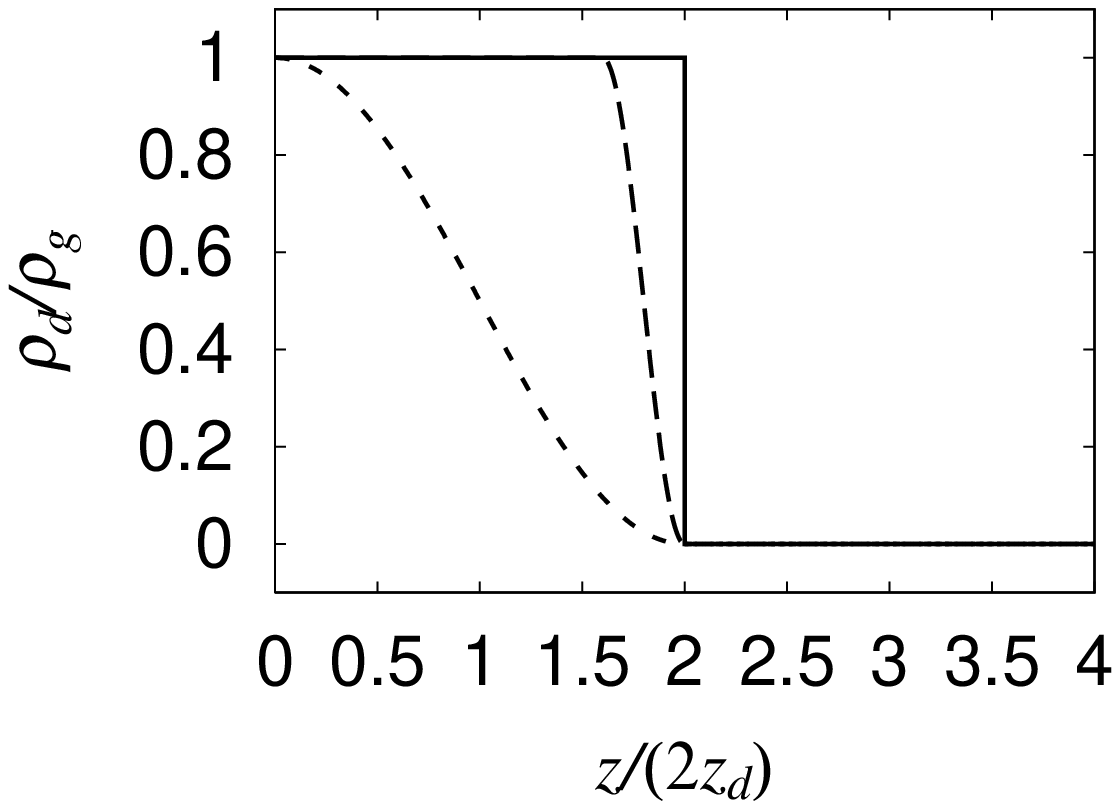}{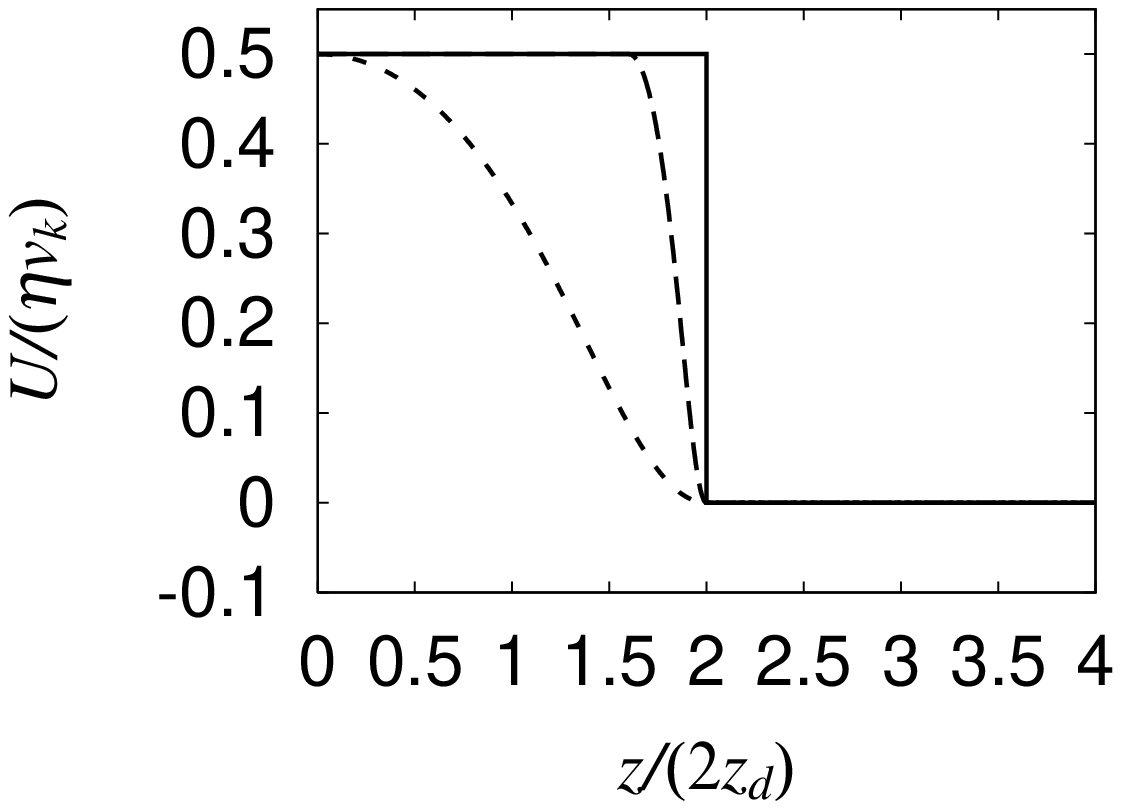}
\caption{The unperturbed density and the unperturbed velocity in the case of $\rho_\mathrm{d}/\rho_\mathrm{g}=1$, $a=1\mathrm{cm}$. The solid line corresponds to step function density profile discussed in Section 3. The dashed line ($h_\mathrm{d}=0.1 z_\mathrm{d}$) and the dotted line ($h_\mathrm{d}=0.5 z_\mathrm{d}$) correspond to continuous density profile.
}
\label{fig:urhohddep.eps}
\end{figure}

\begin{figure}
\plottwo{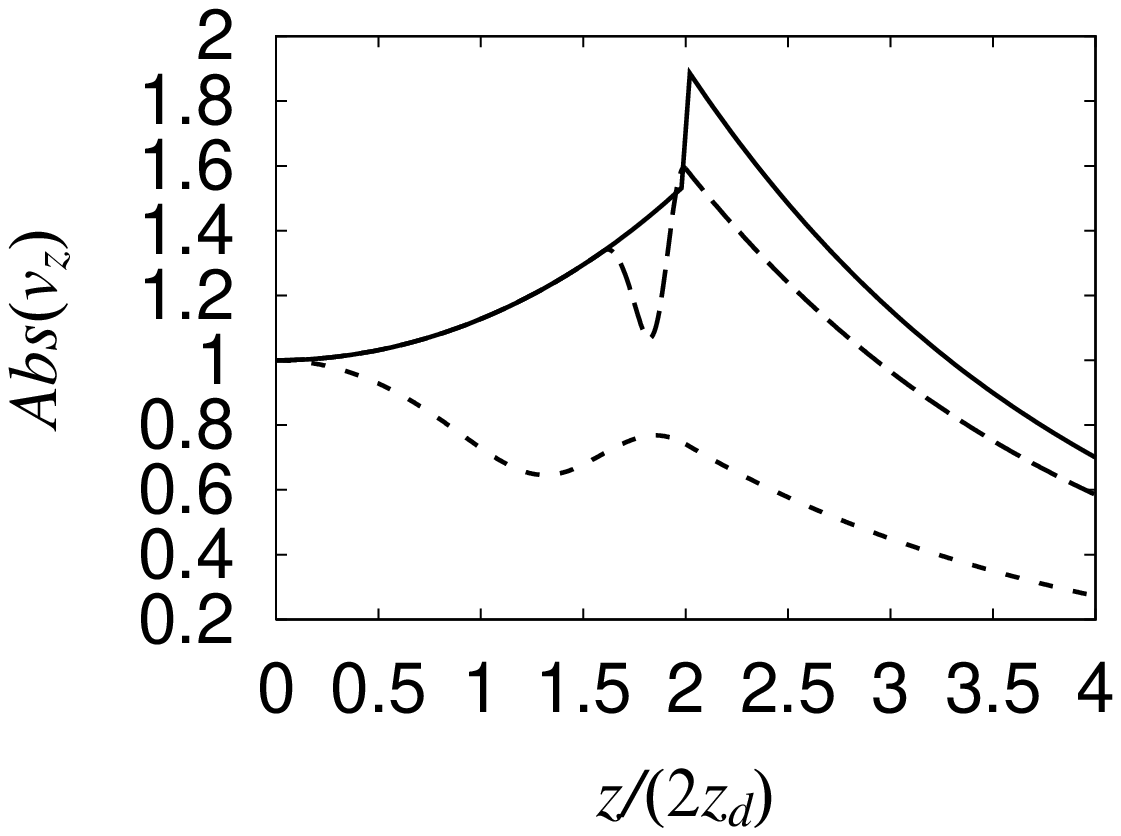}{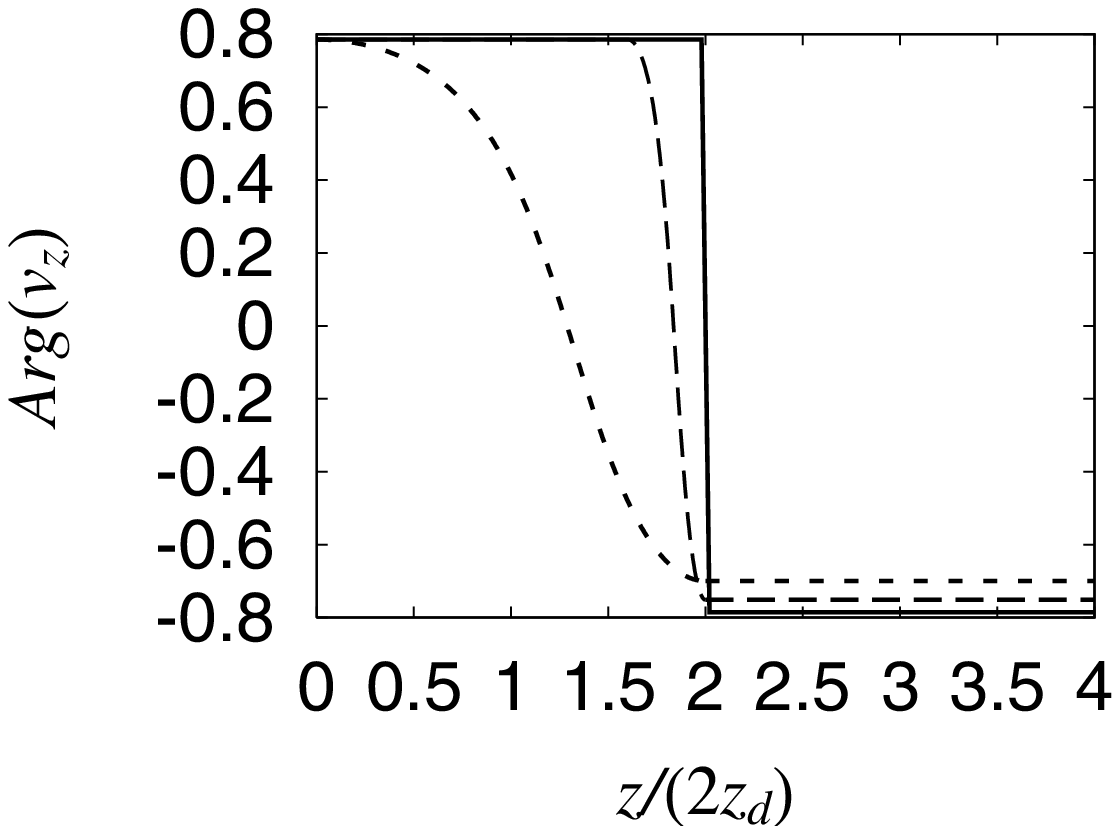}
\caption{Same as Fig. \ref{fig:urhohddep.eps}, but for the eigenfunction of $v_{\mathrm{g}z1}$.}
\label{fig:Eigenhddep.eps}
\end{figure}

\begin{figure}
\plottwo{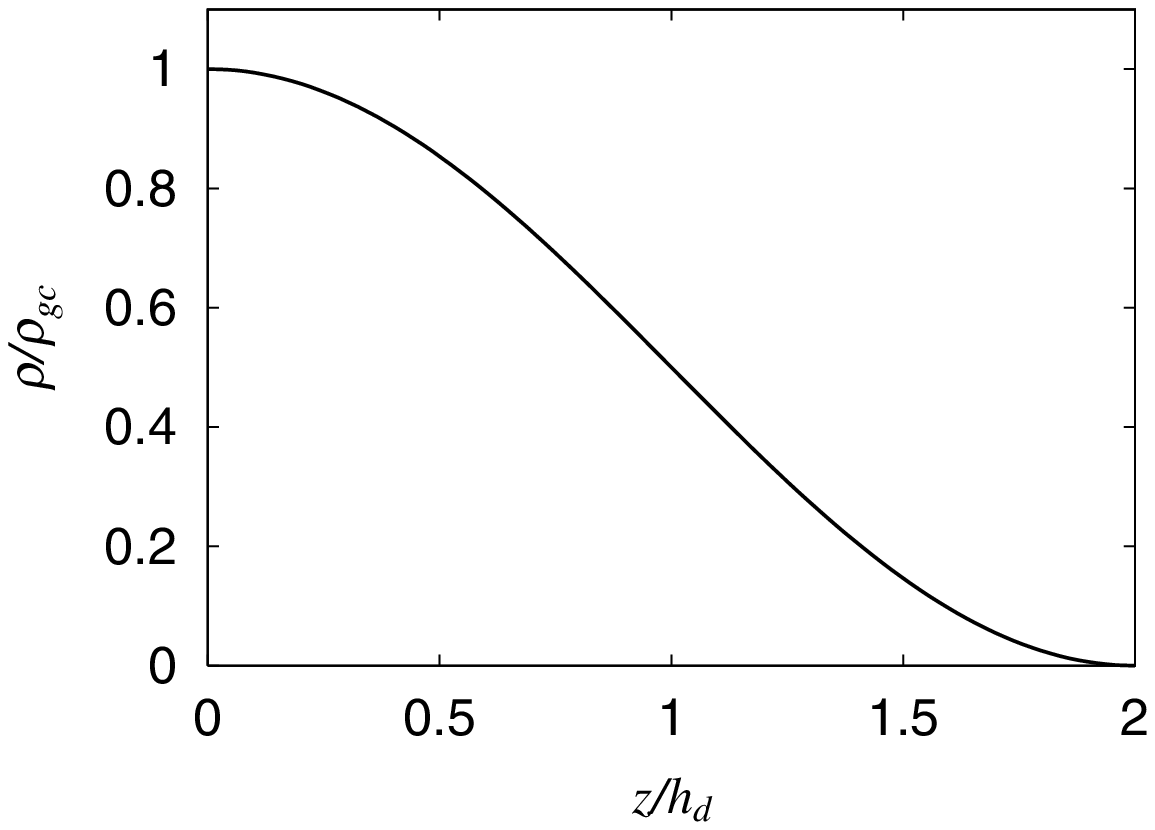}{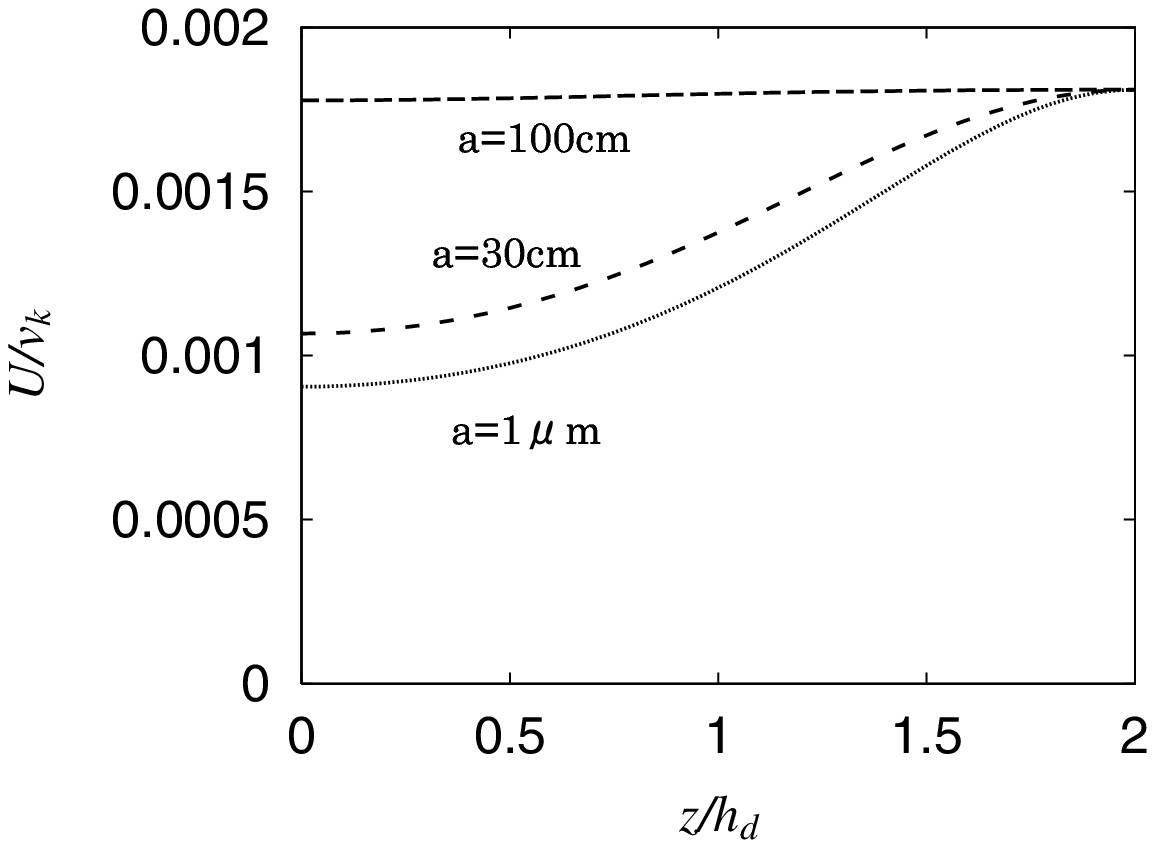}
\caption{The distribution of the unperturbed dust density $\rho_{\mathrm{d}0}$ (\textit{left panel}) and the unperturbed velocity (\textit{right panel}), where $\rho_\mathrm{d}/\rho_\mathrm{g}=1$, $a=1 \mu \rm{m} ,30 \rm{cm} , 100\rm{cm}$, $h_\mathrm{d}=0.5z_\mathrm d$ . We use a frame of reference moving with Kepler velocity.}
\label{fig:RHOandU.eps}
\end{figure}

As shown in Figure \ref{fig:urhohddep.eps}, the hybrid dust density distribution has a constant density region and a transition region whose width is $2 h_\mathrm{d}$. In the limit of $h_\mathrm{d} \to 0$, the dust density distribution becomes the step function discussed in the Section 3. 
In Figure \ref{fig:Eigenhddep.eps}, 
we can see that the eigenfunction of $v_{\mathrm{g}z1}$ approaches to the eigenfunction in the case of the step function. 
We restrict ourselves to the case of $h_\mathrm{d}=0.5 z_\mathrm{d}$ in the following analysis.
The unperturbed state is characterized by two parameters: the size of dust $a$ and the ratio of the dust density to the gas density $\rho_{\rm{d}}/\rho_{\rm{g}}$ on the midplane. Figure \ref{fig:RHOandU.eps} shows the distribution of the unperturbed density and unperturbed velocity.

\begin{figure}
\plotone{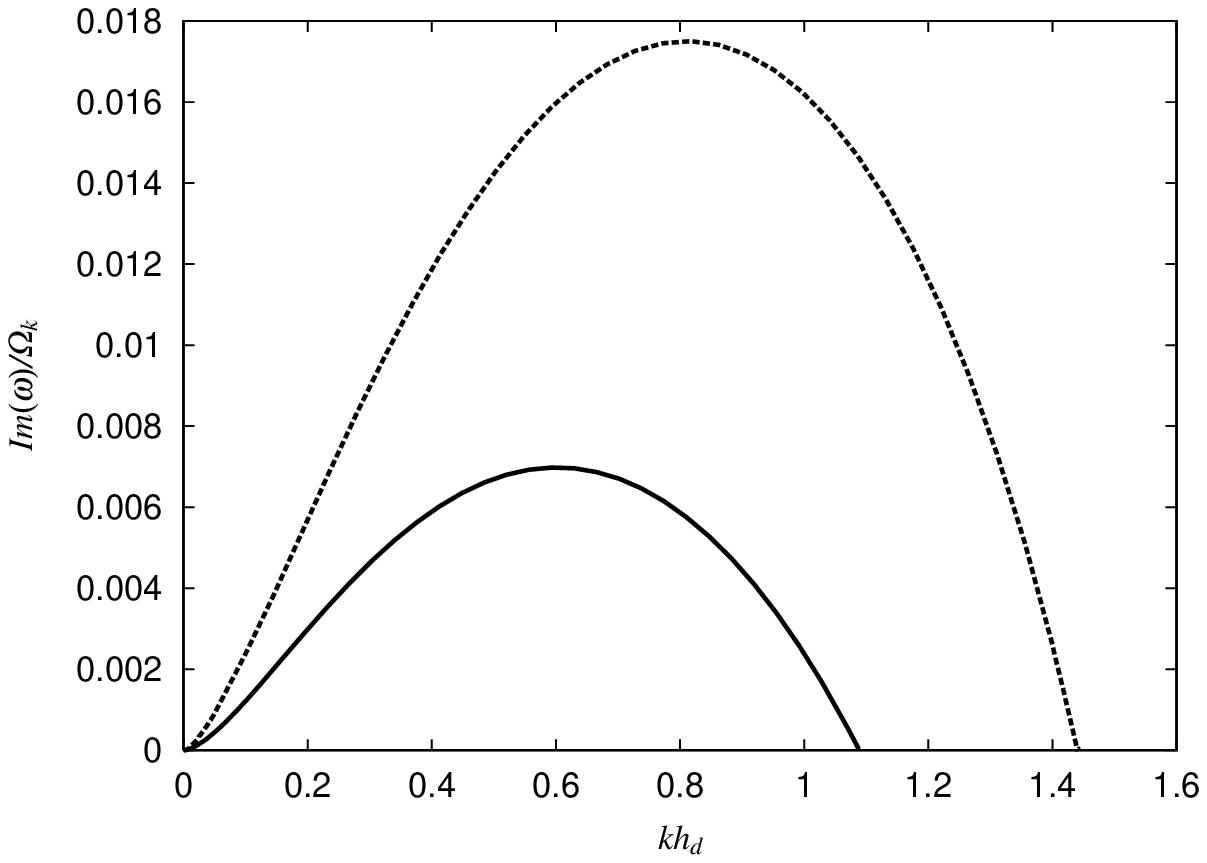}
\caption{The growth rate of the Kelvin-Helmholtz instability as a function of wave number in the case for $a=0.01\rm{cm}$, $\rho_\mathrm{d}/\rho_\mathrm{g}=0.1$, where $h_{\mathrm{d}}=z_{\rm{d}}/2$. The solid line shows the growth rate for the even mode. The dotted line shows the growth rate for the odd mode.}
\label{fig:dispersiondm1sm1.eps}
\end{figure}

Figure \ref{fig:dispersiondm1sm1.eps} shows the growth rate as a function of wave number in the case of $a=0.01\rm{cm}$ and $\rho_{\rm{d}}/\rho_{\rm{g}}=0.1$.
In the limit of $k \to 0$, the growth rate approaches to 0. The growth rate has a peak at a finite wavenumber. 
The modes with short wavelength ($k>k_\mathrm{c}$) are stable.
The odd modes are more unstable than the even modes. 
Thus we restrict ourselves to the odd mode.

\begin{figure}
\plotone{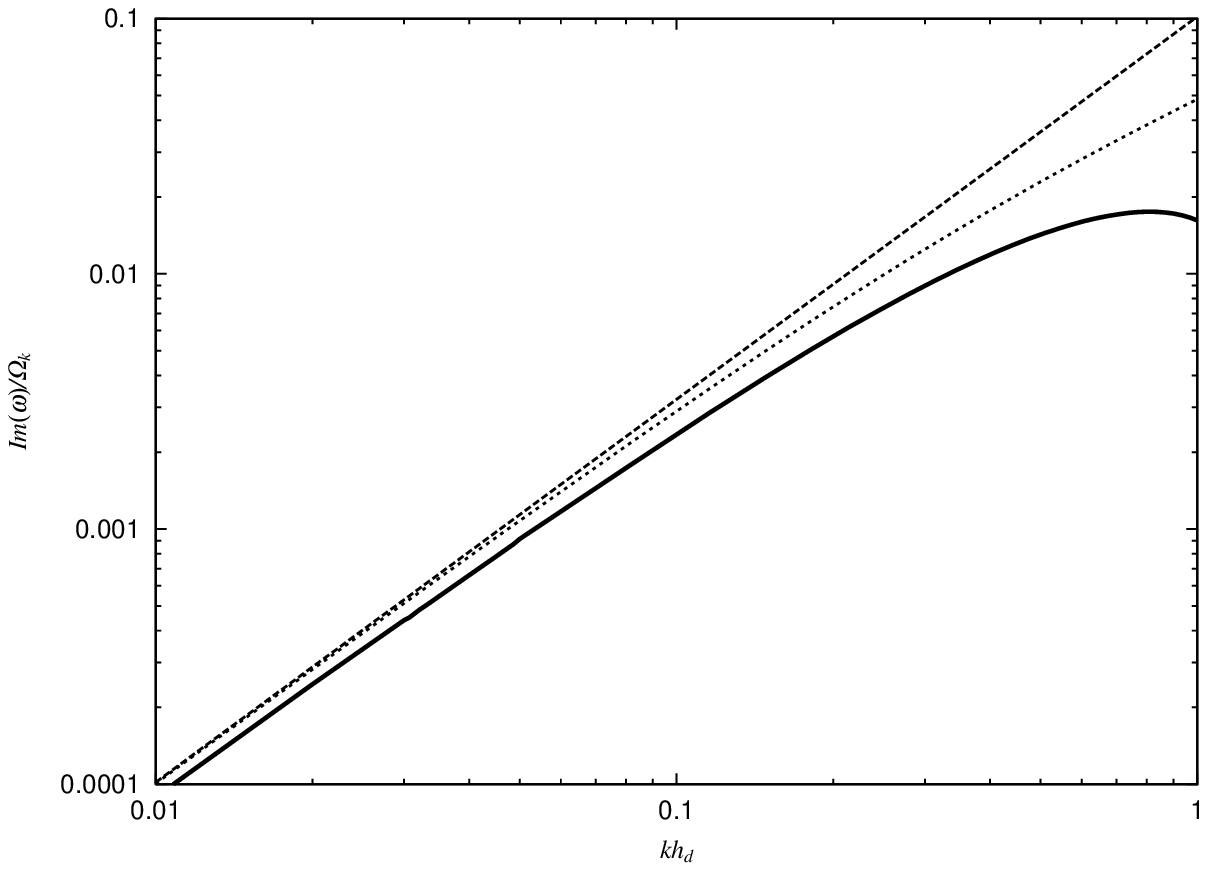}
\caption{Same as Fig.\ref{fig:dispersiondm1sm1.eps}, but in logarithmic scales. The solid line shows the numerical result for the odd mode. The dotted line and the dashed line show the analytic solution (\ref{eq:KH2:SC}) and (\ref{eq:KH2:LW}) respectively.}
\label{fig:dispersion_ana.eps}
\end{figure}

In the limit of $k \to 0$, (i.e., $\lambda = 2 \pi/ k \gg h_{\mathrm{d}} = z_\mathrm{d}/2$), the width of transition region is much smaller than the wavelength, thus we expect that the velocity profile approximately corresponds to the step function discussed in Section \ref{sec:KH2}. As shown in Figure \ref{fig:dispersion_ana.eps}, the growth rate is approximately
proportional to $k^{3/2}$, and approaches to 0 as $k\to0$. 

In the limit of $k \to \infty$, the wavelength is much shorter than the width of transition region. If the unperturbed velocity is continuous, the velocity shear is negligible, hence the flow is stable in this limit.

\begin{figure}
\plottwo{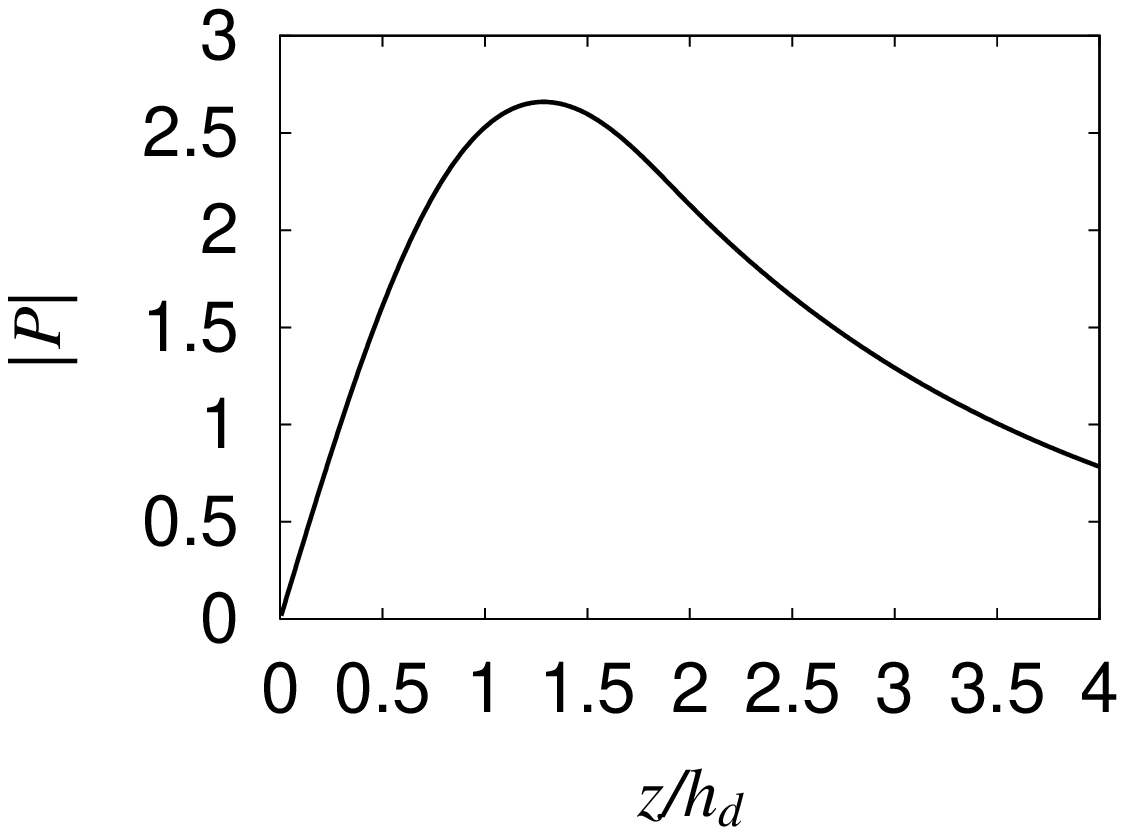}{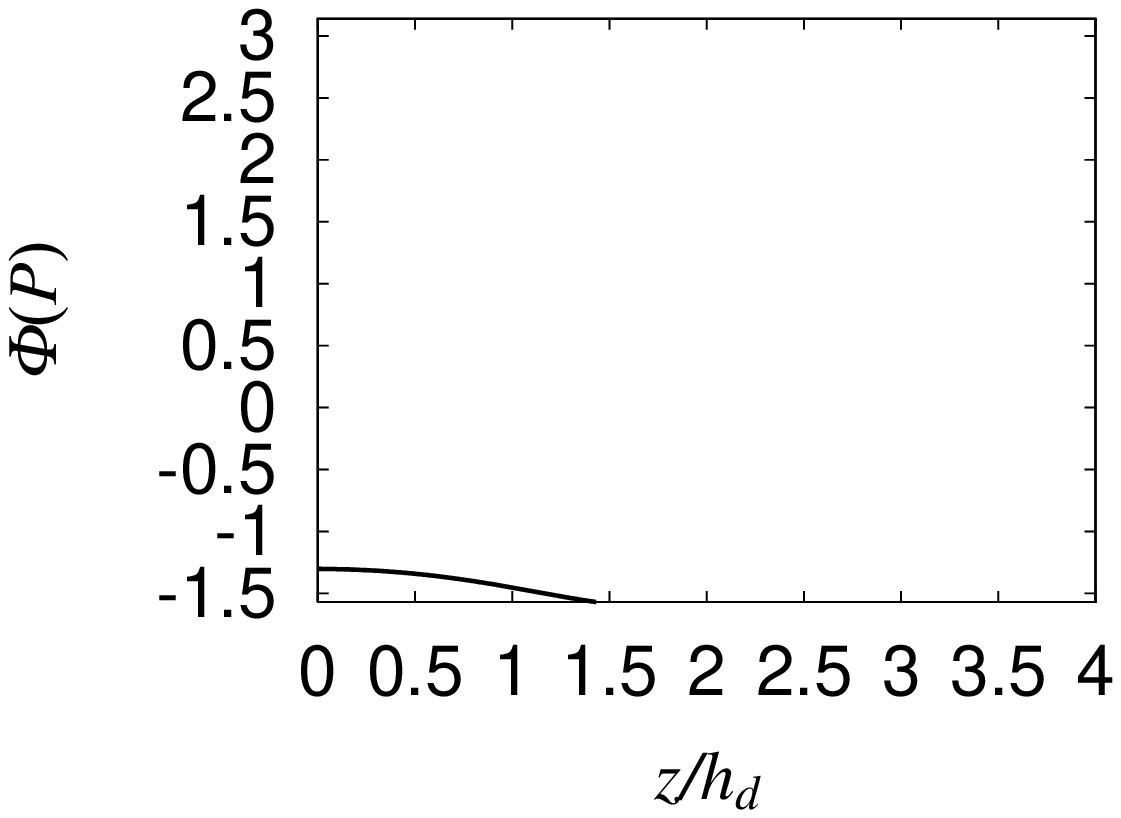}
\caption{
The amplitude (\textit{left panel}) and the phase (\textit{right panel}) of the eigenfunction $P_1(z)$ in the case of $a=1\rm{cm}$ and $\rho_\mathrm{d}/\rho_\mathrm{g}=1$. }
\label{fig:EigenPressureAbsModelA.eps}
\end{figure}

In Figure \ref{fig:EigenPressureAbsModelA.eps},
the amplitude $|P_1(z)|$  and the phase $\Phi(P_1(z))$ are drawn in the case of $a=1\rm{cm}$ and $\rho_{\rm{d}}/\rho_{\rm{g}}=1$, where $\Phi(f)= \tan^{-1} \mathrm{Im} [\mathit{f}]/\mathrm{Re} [\mathit{f}]$.  $|P_1(z)|$ have a maximum value at $z/h_{\rm{d}}\simeq 1.2$, and decreases to 0 as $z/ h_{\rm{d}} \to \infty$. $\Phi(P_1(z))$ varies monotonically in the dust layer $z/h_{\rm{d}}<2$, and become almost constant in region outside the dust layer $z/h_{\rm{d}}>2$.

\begin{figure}
\plottwo{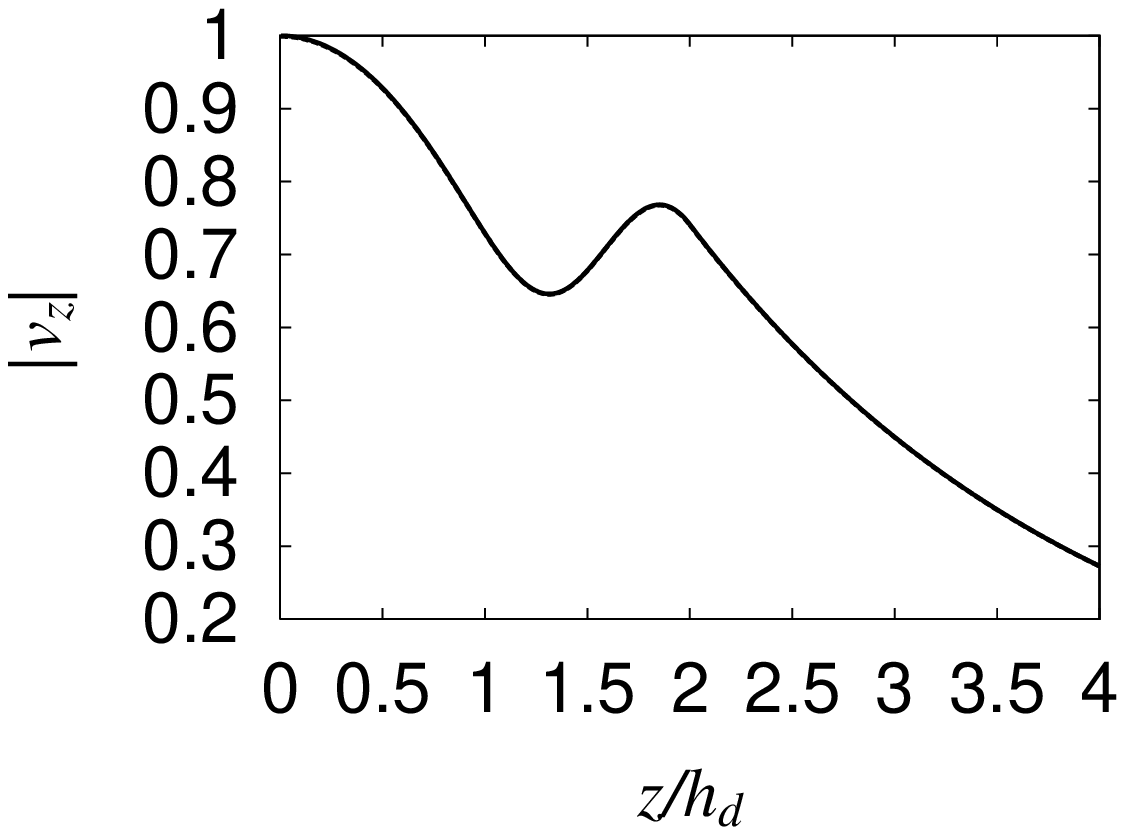}{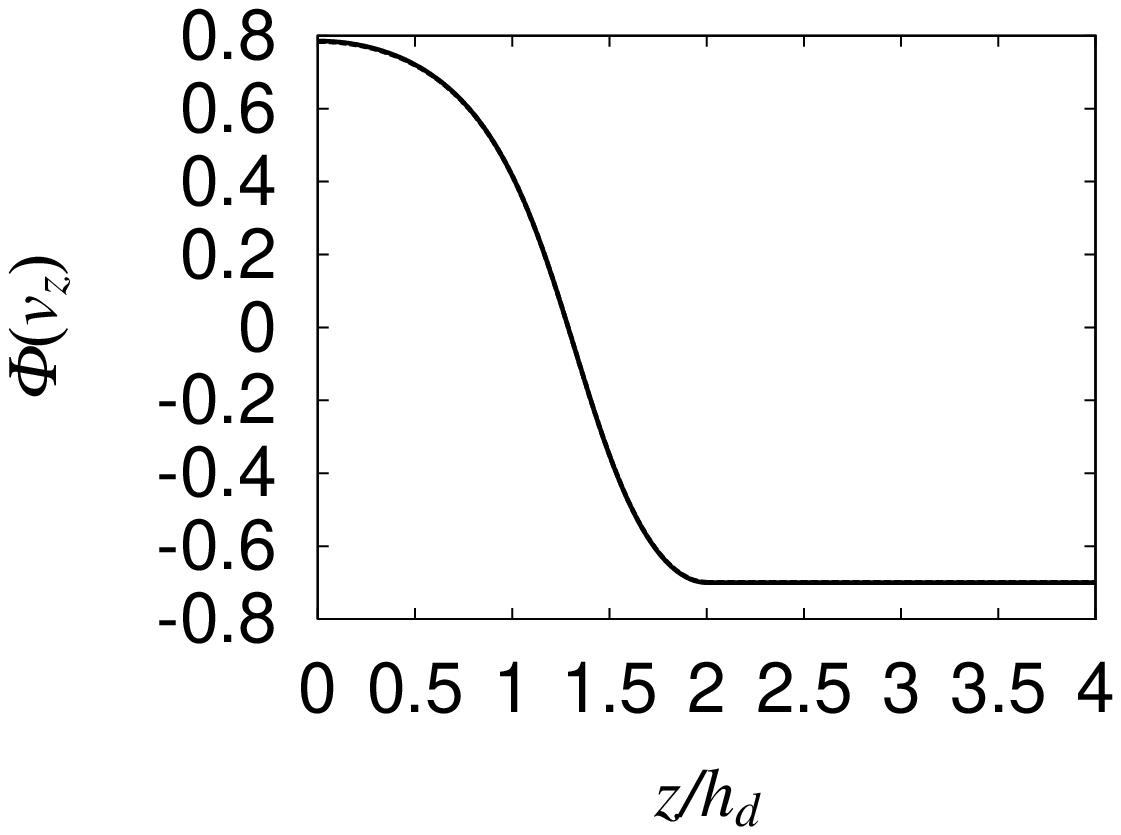}
\caption{The amplitude (\textit{left panel}) and the phase (\textit{right panel}) of eigenfunctions $v_{\rm{g}z1}(z)$ (solid curve) and $v_{\rm{d}z1}(z)$ (dashed curve) in the case for $a=1\rm{cm}$ and $\rho_{\rm{d}}/\rho_{\rm{g}}=1$. The dashed curves are not seen because they overlap on the solid curves.}
\label{fig:EigenVeloZAbsModelA.eps}
\end{figure}

\begin{figure}
\plottwo{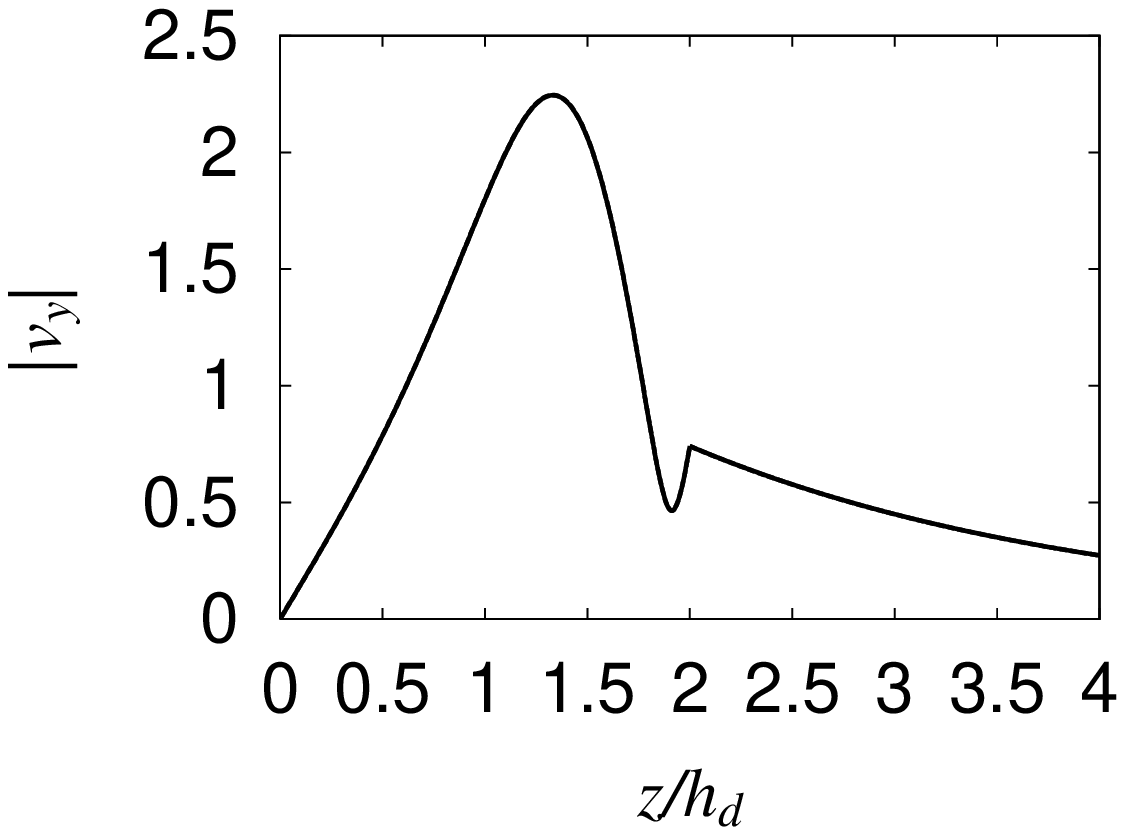}{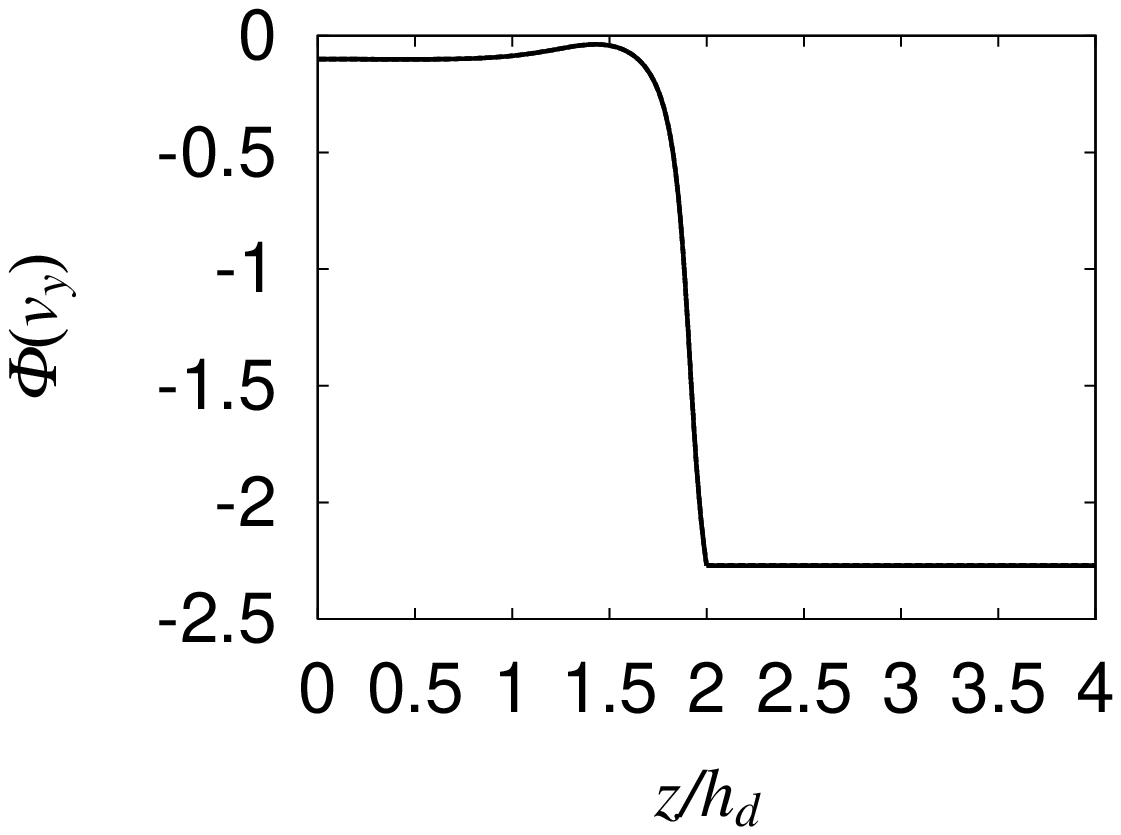}
\caption{Same as Fig.\ref{fig:EigenVeloZAbsModelA.eps}, but for the eigenfunctions $v_{\rm{g}y1}(z)$ (solid curve) and $v_{\rm{d}y1}(z)$ (dashed curve). The dashed curves are not seen because they overlap on the solid curves.}
\label{fig:EigenVeloYAbsModelA.eps}
\end{figure}

\begin{figure}
\plottwo{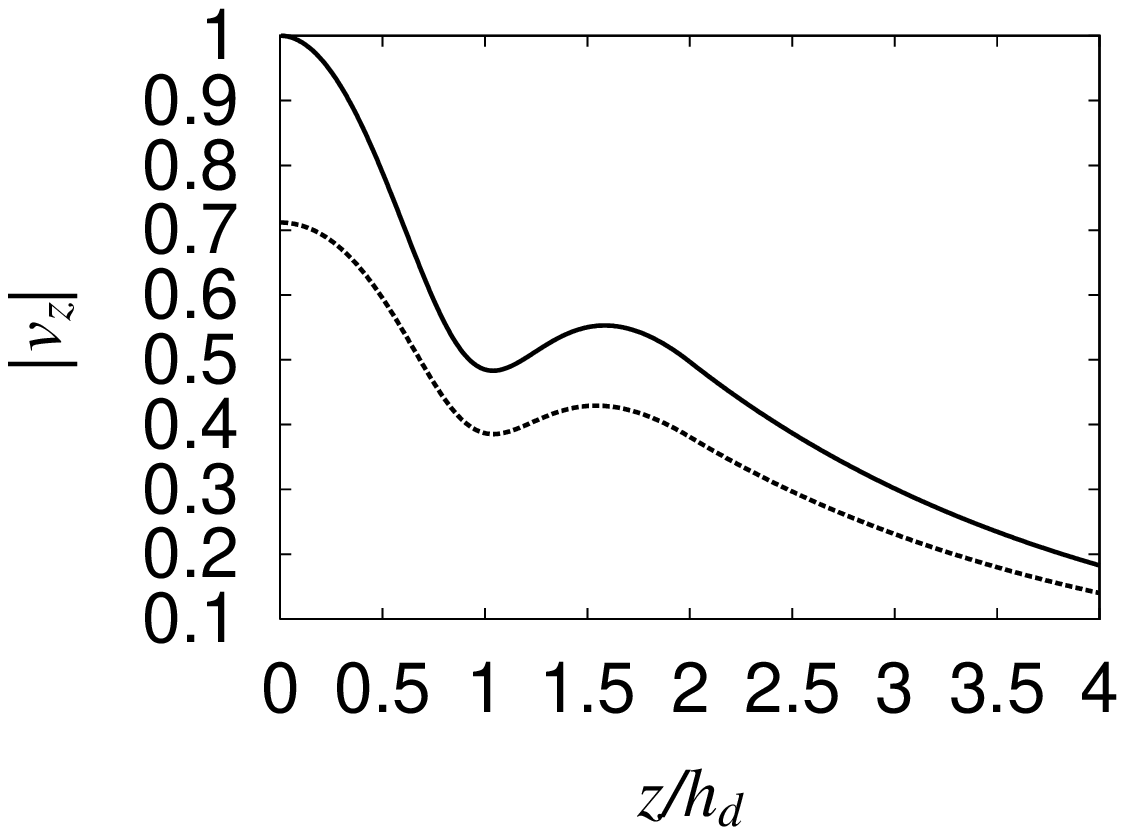}{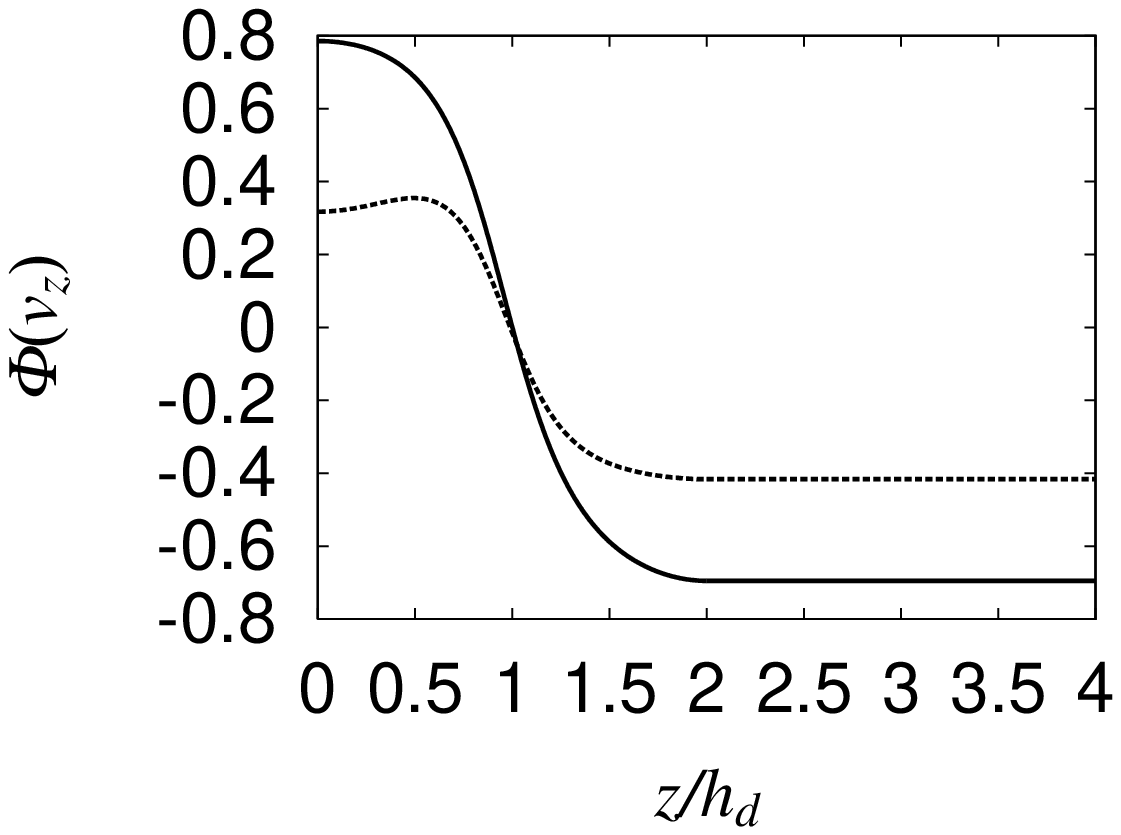}
\caption{Same as Fig.\ref{fig:EigenVeloZAbsModelA.eps}, but for $a=100\rm{cm}$ and $\rho_{\rm{d}}/\rho_{\rm{g}}=1$. }
\label{fig:EigenVeloZAbsModelB.eps}
\end{figure}

\begin{figure}
\plottwo{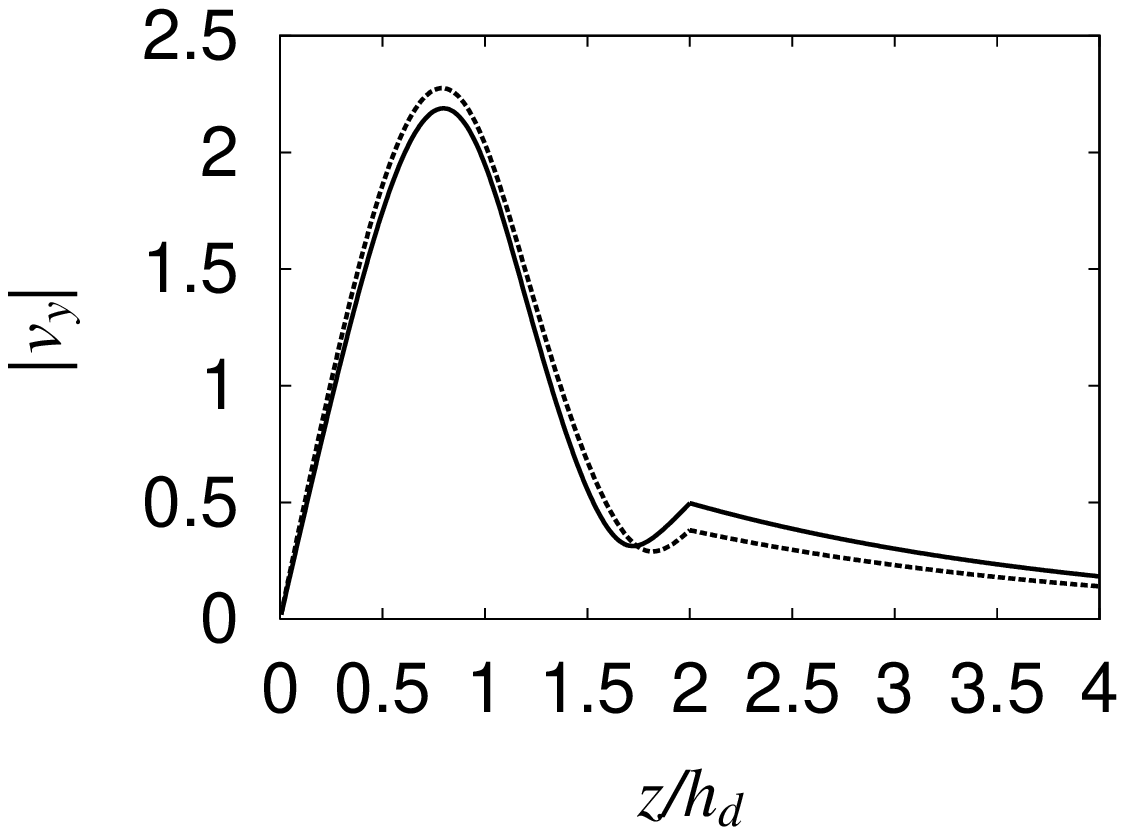}{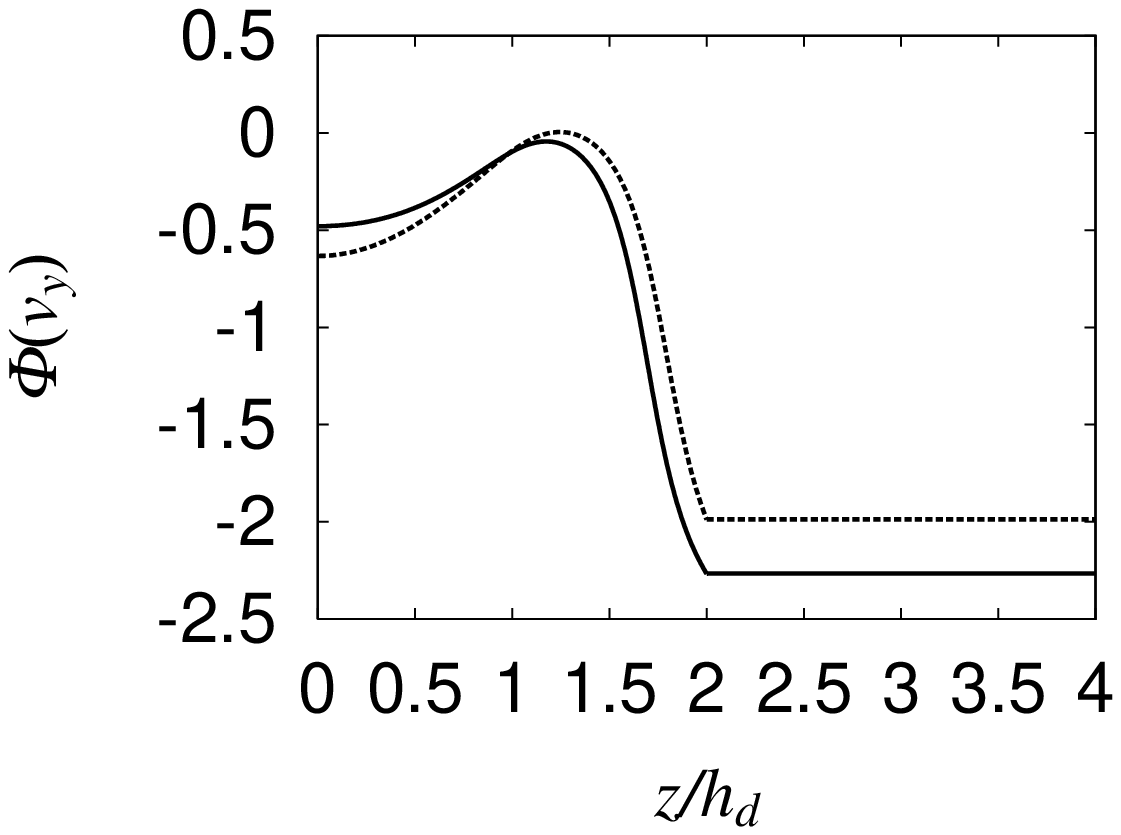}
\caption{Same as Fig.\ref{fig:EigenVeloZAbsModelB.eps}, but for the eigenfunctions $v_{\rm{g}y1}(z)$ (solid curve) and $v_{\rm{d}y1}(z)$ (dashed curve).}
\label{fig:EigenVeloYAbsModelB.eps}
\end{figure}

\begin{figure}
\plotone{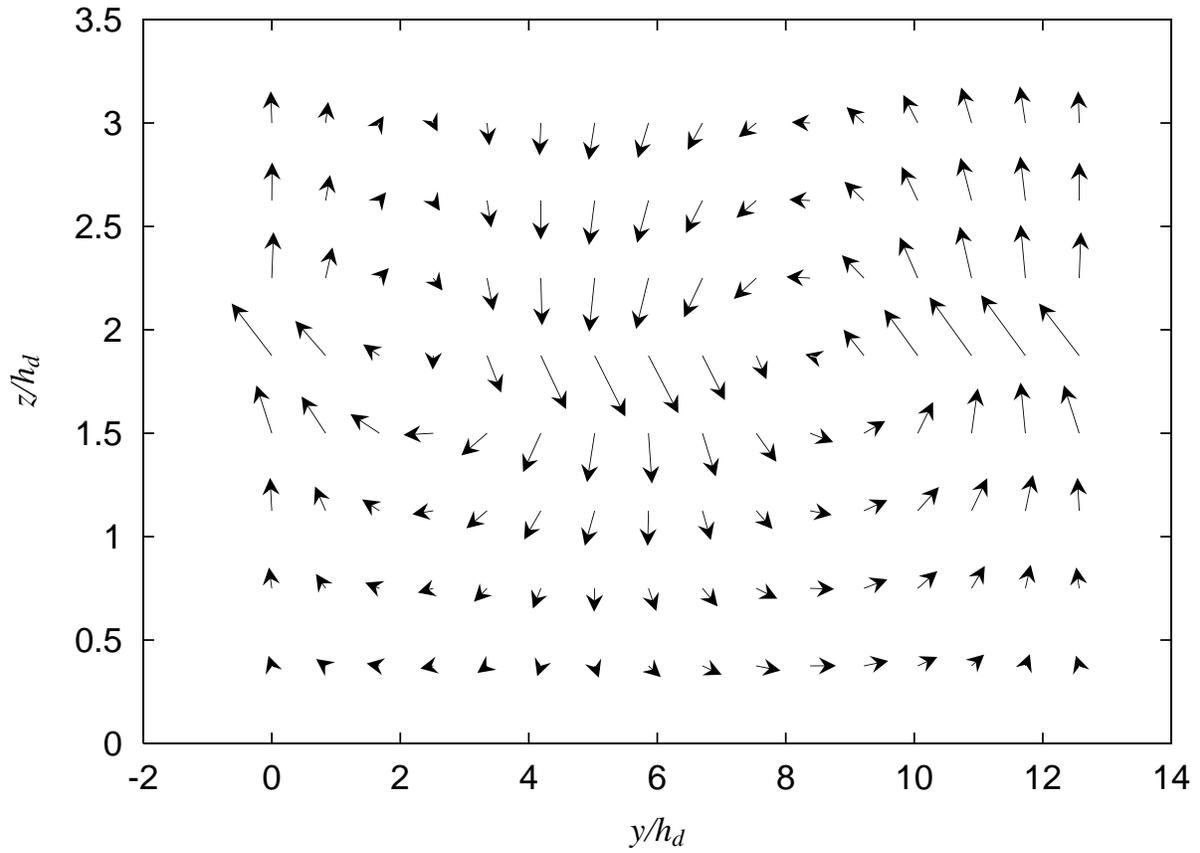}
\caption{The velocity field of gas in the case where $a=1\rm{cm}$ and $\rho_{\rm{d}}/\rho_{\rm{g}}=1$. }
\label{fig:VectorFieldmodelA.eps}
\end{figure}

\begin{figure}
\plotone{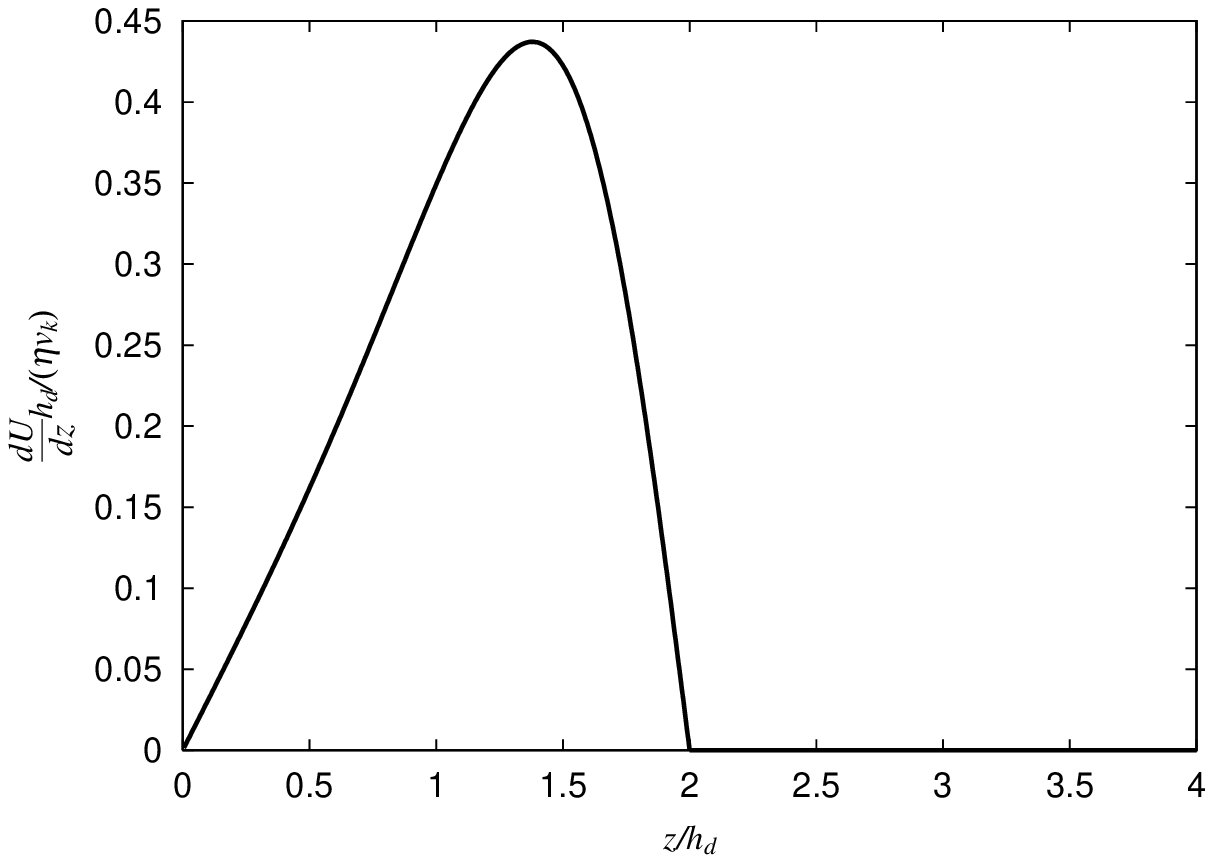}
\caption{The shear rates of the unperturbed state in the case where $a=1\rm{cm}$ and $\rho_{\rm{d}}/\rho_{\rm{g}}=1$.}
\label{fig:ShearRatemodelA.eps}
\end{figure}

In Figure \ref{fig:EigenVeloZAbsModelA.eps} and Figure \ref{fig:EigenVeloYAbsModelA.eps}, the eigenfunctions of velocity are drawn in the case of $a=1\rm{cm}$ and $\rho_{\rm{d}} / \rho_{\rm{g}}= 1$. Since the friction is sufficient to couple the gas and dust firmly, there is almost no difference of velocities between gas and dust. $|v_{\rm{g}z}|$ and $|v_{\rm{g}y}|$ have maximum values at $z/ h_{\rm{d}} \simeq 1.8, 1.3$, respectively.
As shown in Figure \ref{fig:ShearRatemodelA.eps}, the shear rate has a maximum value at $z/ h_{\rm{d}}\simeq 1.4$. 
The velocity field of gas is drawn in Figure \ref{fig:VectorFieldmodelA.eps}. We can see the vortex clearly at $z/h_{\mathrm{d}}=1.5$ which is near the point of the maximum shear rate. The unperturbed quantities interact with the perturbed quantities at the point of the maximum shear rate physically, and the perturbed kinetic energy of the y-component is supplied by the shear (see detailed discussion in \citet{Sekiya2000}).
In Figure \ref{fig:EigenVeloZAbsModelB.eps} and Figure \ref{fig:EigenVeloYAbsModelB.eps},
the eigenfunctions of velocities are drawn in the case of $a=100\rm{cm}$.
The size of dust aggregates is too large for strong coupling, thus the velocity difference is large.
As shown in Figure \ref{fig:EigenVeloYAbsModelA.eps} and Figure \ref{fig:EigenVeloYAbsModelB.eps},
$ d|v_{\rm{g}y1}|/dz$ is discontinuous at $z/h_{\rm{d}}=2$, since there is no boundary condition that $|v_{\rm{g}y1}|$ is differentiable. On the other hand, in Figure \ref{fig:EigenVeloZAbsModelA.eps}, $|v_{\rm{g}z1}|$ is differentiable owing to two boundary conditions at $z=2h_{\rm{d}}$: $v_{\mathrm{g}z1}/(\omega-k U)$ and the pressure perturbation are continuous.

\begin{figure}
\plotone{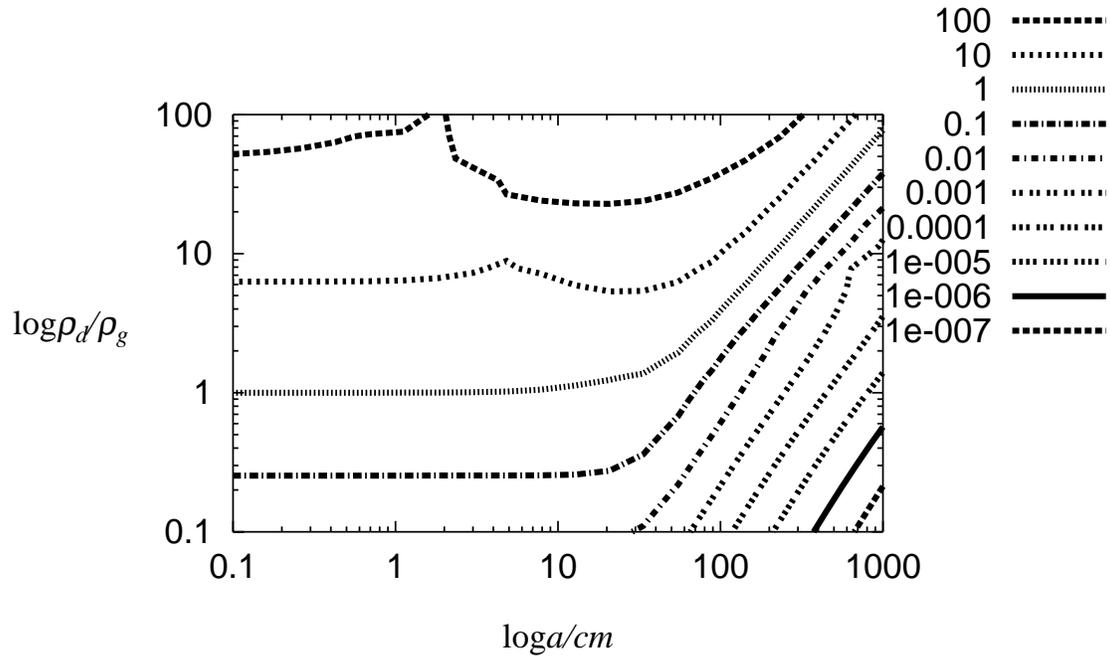}
\caption{The contour of the peak growth rate of the Kelvin-Helmholtz instability as function of the dust to gas density ratio on the midplane and the size of dust. The growth rate is normalized by the Keplerian frequency.}
\label{fig:contour(maximum)fd1Ub.eps}
\end{figure}

The growth rate of the mode with the most unstable wave number as functions of the dust to gas density ratio on the midplane and the size of dust aggregate is drawn in 
Figure \ref{fig:contour(maximum)fd1Ub.eps}.
The growth rate increases as $\rho_{\rm{d}}/\rho_{\rm{g}}$ increases. If $\rho_{\rm{d}}$ is large, the thickness of the dust layer $h_{\rm{d}} \simeq \Sigma_{\rm{d}} / \rho_{\rm{d}}$ is thin, and the velocity difference is large, thus the shear becomes large. Hence the dust layer becomes more unstable as the dust settling proceeds.
When $\rho_{\rm{d}}/\rho_{\rm{g}}>1$, under the condition that $\rho_{\rm{d}}/\rho_{\rm{g}}$ is constant,
the growth rate has the local minimum at $a\sim 1\rm{cm}$. When $a>100\rm{cm}$, the growth rate decreases significantly.
The reason of the reduction of the growth rate at $a\sim 1\rm{cm}$ is caused by the friction discussed in Section \ref{sec:KH1}.
However the reduced growth rate is much larger than the Kepler frequency, and the shear instability is induced, dust aggregates are stirred up by turbulence. The reduction in the case of $a>100\rm{cm}$ is not caused by the friction. The friction is so small that the unperturbed velocity is almost constant. Thus the shear is not so large to induce the Kelvin-Helmholtz instability.

We compare the numerical result and the analytic result discussed in Section \ref{sec:KH1}.
We can expect that the characteristic length scale $L$ is the thickness of dust layer. On the other hand, the analysis in Section \ref{sec:KH1} has no characteristic length scale.
If we introduce the characteristic length scale into the analysis in Section \ref{sec:KH1},
we may compare the result in this Section with the result in Section \ref{sec:KH1}. 
Thus we consider  $k$ as $2 \pi/L$. When the stabilization due to the inertia of dust is not effective, the maximum growth rate may be expressed as $2 \pi U/L$ where $U$ is the characteristic velocity difference.

The velocity difference between the midplane and the gas dominant region is expressed as the equation (\ref{eq:steady_velo}).
In the small friction limit ($A ( \rho_\mathrm{g} + \rho_\mathrm{d}) \ll \Omega_k$), we have
\begin{equation}
\delta v_{\mathrm{g}} \simeq \mathcal{R}_\mathrm{d}(1+\mathcal{R}_\mathrm{d})\Gamma^2 \eta v_k.
\end{equation}
In the strong friction limit, we have
\begin{equation}
\delta v_{\mathrm{g}}=\frac{\mathcal{R}_\mathrm{d}}{\mathcal{R}_\mathrm{d}+1} \eta v_{\rm{k}}.
\label{eq:scvelo}
\end{equation}
The thickness of the dust layer is
\begin{equation}
z_\mathrm{d} \sim \Sigma_{\rm{d}} / \rho_{\rm{d}}.
\label{eq:heightex}
\end{equation}

With (\ref{eq:condstab}), (\ref{eq:scvelo}), and (\ref{eq:heightex}), we obtain the condition of stabilization:
\begin{equation}
\frac{20}{\mathcal{R}_\mathrm{d}} \rm{cm} < a < 10\rm{cm},
\label{eq:stablecondition}
\end{equation}
where we assume $\mathcal{R}_\mathrm{d} \gg 1$ and $\mu=0.5$. 
As shown in Figure \ref{fig:contour(maximum)fd1Ub.eps}, the region where the growth rate is reduced owing to the friction corresponds to the condition (\ref{eq:stablecondition}), thus the numerical result is consistent with the condition discussed in Section \ref{sec:KH1}.

In the case of $a\gg 1\rm{cm}$, the friction is very weak, and the reduction of growth rate due to friction is not effective. The friction coefficient $A$ is proportional to $a^{-2}$ in the Stokes regime.
So we can estimate the growth rate $\mu$:
\begin{equation}
\mu \propto a^{-4}\mathcal{R}_\mathrm{d}^2(1+\mathcal{R}_\mathrm{d})
\end{equation}
Thus the maximum growth rate is inversely proportional to $a^4$ in the case of $a\gg 1\rm{cm}$, and proportional to $\mathcal{R}_\mathrm{d}^2$ in the case of $\mathcal{R}_\mathrm{d} < 1$, or proportional to $\mathcal{R}_\mathrm{d}^3$ in the case of $\mathcal{R}_\mathrm{d} > 1$.

\clearpage

\section{Discussion \label{sec:discussion}}
When the size of the dust grain $a$ is smaller than $1\rm{cm}$, the stabilization by the friction is not effective. Thus the dust layer is expected to become turbulent, if the settling onto the midplane proceeds. When $a>100\rm{cm}$, however, the growth rate of Kelvin-Helmholtz instability is small, the settling may proceed without stirring up. In this section we discuss a possible path to the planetesimal formation through the gravitational instability.

We compare the various timescales for dust grains in the disk as functions of the size of dust and the degree of the dust settling onto the midplane.
For simplicity, we assume that all dust aggregates have the same size.
We consider the timescales of the dust settling, Kelvin-Helmholtz instability, the dust growth, and the migration into the central star.

The dust growth time can be calculated by
\begin{equation}
t_{\rm{grow}}=\frac{4}{f_{\rm{s}}} \frac{\rho_{\rm{mat}}}{\rho_{\rm{d}}} \frac{a}{\delta v},
\end{equation}
where $f_s$ is sticking probability, $\rho_{\rm{mat}}$ matter density, and $\delta v$ the averaged velocity dispersion of the dust aggregates.
$\delta v$ is given by 
\begin{equation}
\delta v=\max(\delta v_1,\delta v_2),
\end{equation}
where 
\begin{equation}
\delta v_{1}= \sqrt{\frac{m_H}{m_d}} v_{gt},
\end{equation}
\begin{equation}
\delta v_{2}= \frac{\Omega_{\rm{k}}}{A \rho_{\rm{g}}} \frac{z}{r} v_{\rm{k}},
\label{eq:relvelo}
\end{equation}
$m_{\rm{H}}$ hydrogen mass, $m_{\rm{d}}$ dust mass, $z=z_{\rm{d}}$, $v_{\rm{gt}}$ the thermal velocity of gas.

In the laminar flow, we use the thermal velocity $\delta v=\delta v_1$ given by the energy equipartition of gas and dust.
This is expected if the dust aggregate is small.
If the dust aggregates are large and settling, the dust aggregates collide with the other small dust aggregates which are not settling. In this case the relative velocity $\delta v$ might be determined by the settling velocity $\delta v_2$.

The sticking probability is nearly unity in the case of the collision with small relative velocity. In the case of large relative velocity, the sticking probability decreases to zero \citep{Poppe2000,Sekiya2003}. However we use the sticking probability $f_s=1$ for simplicity, because the smaller sticking probability should not change the following diagram.

In addition we use the equation (\ref{eq:relvelo}) for large dust aggregates.
In the laminar flow the final sizes of dust aggregates are $20 \mathrm{cm}$ at 1AU \citep{Nakagawa1986}.
In the case of large dust aggregates, we can not use the equation (\ref{eq:relvelo}) to calculate the time scale of the growth of dust aggregates. We can use the equation (\ref{eq:relvelo}) for small dust aggregates. 
On the other hand, if the dust layer is turbulent the equation (\ref{eq:relvelo}) cannot be used.  
However we are interested in the boundary line between turbulent dominant state and settling dominant state. This boundary state does not depend on the growth time if the growth time of the dust aggregates is longer than the settling time. 
The settling for large dust aggregates is very fast, thus we can expect that the settling time is shorter than the time scale of growth in the case of large dust aggregates. Thus we use the equation (\ref{eq:relvelo}) for simplicity. The boundary line should not change even if we use more realistic growth time. 

If a dust aggregates is small ($a<20\mathrm{cm}$), the settling time onto the midplane is determined by 
\begin{equation}
t_{\rm{settle}} = \frac{A \rho_{\rm{g}} }{\Omega_{\rm{k}}^2}.
\end{equation}
However if a dust aggregates is large ($a>20\mathrm{cm}$), it revolves around the central star decreasing the inclination gradually \citep{Nakagawa1986}. The decreasing timescale of inclination is order of stopping time $1/A\rho_\mathrm{g}$. Thus in the case of $a>20\mathrm{cm}$, the effective settling time is considered as decreasing timescale:
\begin{equation}
t_{\rm{settle}}=1/A \rho_{\mathrm{g}}.
\end{equation}

The timescale of Kelvin-Helmholtz instability is 
\begin{equation}
t_{\rm{KH}} = \frac{1}{\rm{Im} [\omega_{\rm{max}}]}.
\end{equation}

\begin{figure}
\plotone{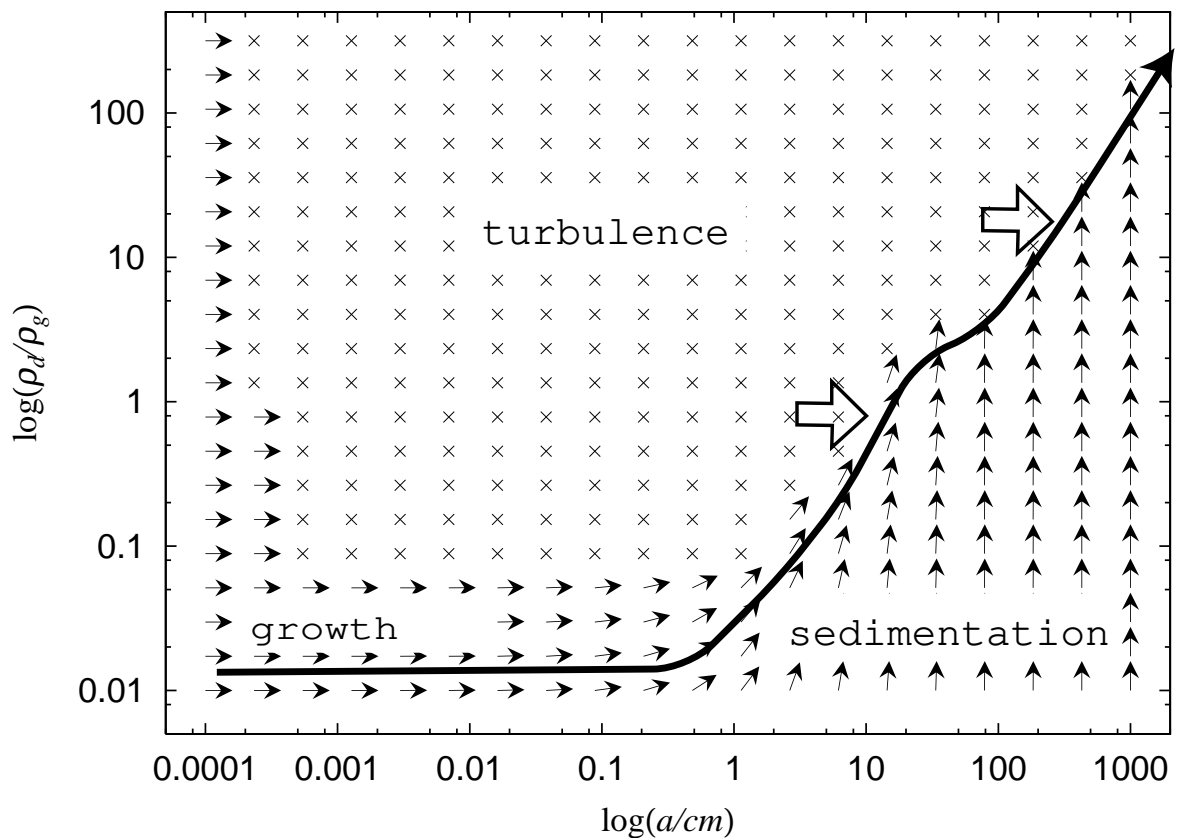}
\caption{The dominant processes in the disk of various parameters. The Kelvin-Helmholtz instability region is denoted by the crosses. The thin arrows mean the variation of $a$ and $\rho_\mathrm{d}/\rho_\mathrm{g}$ in the region where the sedimentation or the growth is dominant process. 
The block arrows mean the growth in the turbulent flow. 
The bold arrow is the possible path to the planetesimal formation through the gravitational instability.}
\label{fig:path_ordinary2.eps}
\end{figure}

The result is summarized in Figure \ref{fig:path_ordinary2.eps}.
In the region of $a<0.1\rm{cm}$ and $\mathcal{R}_\mathrm{d}<0.01$, the growth of dust aggregates is the fastest process, thus the dust can grow up to $a\sim 0.1\rm{cm}$. In the region of $a<0.1\rm{cm}$ and $\mathcal{R}_\mathrm{d}>0.01$, the Kelvin-Helmholtz instability is the fastest process.
In the region of $a>0.1\rm{cm}$, the dust settling is faster than the dust growth, thus we should compare two processes: the settling and the Kelvin-Helmholtz instability. In the region of $a\gg 0.1\rm{cm}$, the settling velocity is very fast and the Kelvin-Helmholtz instability is not effective, thus the settling region expands into higher $\rho_\mathrm{d} /\rho_\mathrm{g}$ region as the size of dust $a$ increases.

We assume that the turbulent flow become laminar if the timescale of Kelvin-Helmholtz instability is larger than 
those of other processes.
If the density and the size of dust aggregates are $\rho_\mathrm{d}/\rho_\mathrm{g} < 0.01, a=10^{-4} \rm{cm}$,
 the dust aggregates grow up to $a\sim 0.1\rm{cm}$ without the dust settling.
When $a > 0.1 \rm{cm}$, the dust aggregates start settling and the Kelvin-Helmholtz instability occurs, which develops turbulence in the dust layer.
The dust aggregates may grow owing to the random velocity field in the turbulent flow, thus the timescale of the Kelvin-Helmholtz instability will increase with the growth of dust. When the size of dust grain becomes sufficiently large, $t_{\rm{KH}} > t_{\rm{settle}}$, the dust aggregates begin to settle. Hence the dust layer may evolve along the boundary line between the turbulence region and the settling region in Figure \ref{fig:path_ordinary2.eps}.
The density of dust in the disk may reach $100 \rho_{\mathrm{g}}$ at $a\sim 10\rm{m}$.

$\rho_{\mathrm{d}}=100 \rho_{\mathrm{g}}$ corresponds to the order of the critical density for the gravitational instability.
However this argument is not conclusive, because the critical density may be larger than $100 \rho_\mathrm{g}$. A more realistic critical density depends on a unperturbed dust density profile. Actually \citet{Yamoto2004} conclude that the critical density for the constant Richardson number dust density distribution is $760\rho_\mathrm{g}$ and that for the Gaussian dust density distribution is $340 \rho_\mathrm{g}$. Thus the realistic condition of the gravitational instability seems to be depend on the circumstances.

\begin{figure}
\plotone{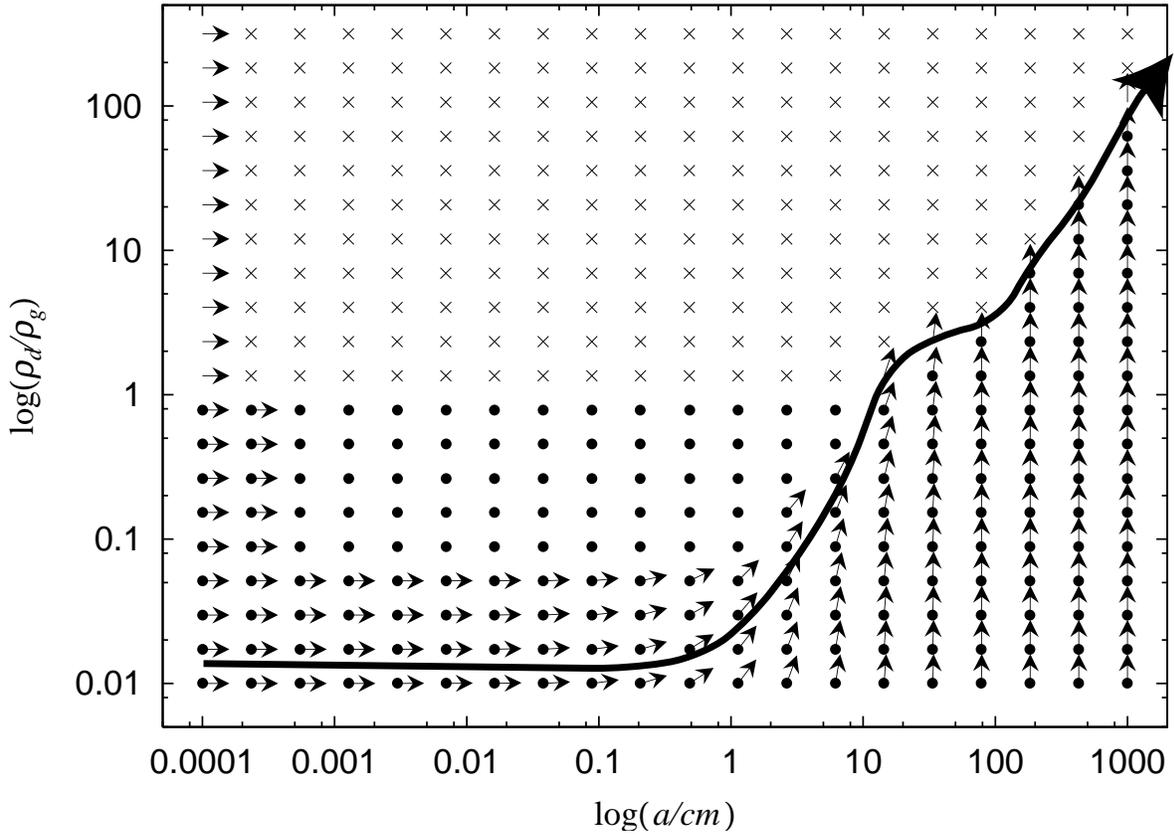}
\caption{Same as Figure \ref{fig:path_ordinary2.eps} but we take into account the stabilizing effect of the Coriolis and tidal forces. Filled circles correspond to the models that are expected to be stabilized by Coriolis and tidal forces.}
\label{fig:path3.eps}
\end{figure}

\citet{Ishitsu2003} showed that the shear instability which grows slower than the Kepler timescale without the Coriolis and tidal forces is suppressed if these forces are taken into account. Their analysis is based on the strong coupling approximation but our analysis is not based on it, thus we cannot use their result directly. To take into account the Coriolis and tidal forces, we assume that the region $t_{\mathrm{KH}} > 2\pi /\Omega_{\mathrm{k}}$ should correspond to a stable region, where  $t_{\mathrm{KH}}$ is the timescale of the shear instability without the Coriolis and tidal forces, $2\pi /\Omega_{\mathrm{k}}$ is the Kepler timescale.
We plot the stable region in Figure \ref{fig:path3.eps}. This effect causes little change on a possible path.

The growth timescale for meter-size particles is longer than the timescale of the migration into central star because of the large radial velocity\citep{Adachi1976,Weidenschilling1977}.
This is correct when the dust density is smaller than the gas density ($\mathcal{R}_\mathrm{d}<1$). The radial velocity in the case of $\mathcal{R}_\mathrm{d}>1$ is slower than the radial velocity in the case of $\mathcal{R}_\mathrm{d}<1$ \citep{Nakagawa1986}.
The radial velocity of dust is expressed as 
\begin{equation}
v_{\mathrm{d}r}=- \frac{2 \Gamma}{1+(1+\mathcal{R}_\mathrm{d})^2\Gamma^2} \eta v_k.
\end{equation}
Thus the migration timescale $t_{\mathrm{migration}}=r/|v_{\mathrm{d}r}|$ is 
\begin{equation}
t_{\mathrm{migration}}=\frac{(1+(1+\mathcal{R}_\mathrm{d})^2\Gamma^2)r}{2 \Gamma \eta v_k}.
\end{equation}
In the limit of $\mathcal{R}_\mathrm{d} \gg 1/\Gamma$, the migration timescale is proportional to $(1+\mathcal{R}_\mathrm{d})^2$.
In the case where $\mathcal{R}_\mathrm{d}=0.01$ and $a=100\mathrm{cm}$, the migration timescale is about $10^2$ year at 1 AU. In contrast, if $\mathcal{R}_\mathrm{d} = 100$ and $a=100\rm{cm}$, the migration timescale is about $10^5$ year. Thus the dust aggregates can avoid migration onto their central star and lead to the gravitational instability, if the growth timescale in the turbulent flow is shorter than the migration timescale $10^5$ year.

\clearpage

\section{Conclusion \label{sec:conclusion}}
In this paper, we have investigated the Kelvin-Helmholtz instability for the dusty fluid taking into account of the effect of relative motion between gas and dust.

 To clarify the physics of Kelvin-Helmholtz instability in the dusty fluid, we have analyzed the linear stability of the flow whose velocity profile is the simple Heaviside step function.
We have found the stabilizing effect due to friction.
In the case of $\rho_{\rm{d}}/\rho_{\rm{g}} \gg 1$, the reduced growth rate is 
\begin{equation}
\frac{\rm{Im} [\omega]}{\rm{Im} [\omega_{\rm{sc}}]}  \simeq \left( \rho_{\rm{d}}/\rho_{\rm{g}} \right)^{-1/2},
\end{equation}
where $\rm{Im} [\omega_{\rm{sc}}]$ is the growth rate of the Kelvin-Helmholtz instability for simple fluid without dust aggregates.

We have considered the case of dusty layer between two gaseous layers. 
Because of the symmetry, there are the even mode ($P_1(z)=P_1(-z)$) and the odd mode ($P_1(z)=-P_1(-z)$).
We have found that the growth rate for odd modes is more unstable than that for even modes.

We have analyzed the linear stability in a more realistic protoplanetary disk where the dust density distribution is a sinusoidal density distribution.
In the case of $a<0.1\rm{cm}$, we found the stabilization due to the friction.
This result is consistent with the result in the case of the step function velocity profile.
However the timescale of the stabilized Kelvin-Helmholtz instability is shorter than that of settling, thus the dust layer becomes turbulent before the gravitational instability sets in.
In the case of $a \gg 1 \rm{cm}$, since the shear is small, the Kelvin-Helmholtz instability is stabilized sufficiently.
In this case, if the dust aggregates begin to settles, it is expected that the dust settles without stirring up and may reach the critical density.

We have compared the timescales of the dust growth, the dust settling, and Kelvin-Helmholtz instability, and proposed a scenario in which the dust layer proceeds along the boundary between the turbulent region and settling region in Figure \ref{fig:path_ordinary2.eps}, and the dust layer reaches the order of critical density of gravitational instability at $a\sim 10\rm{m}$. In this way, the planetesimals may be formed through gravitational instability.

To justify this scenario, we have to do the detailed numerical calculation of the growth of dust aggregates in the turbulent flow.
In this paper, we neglected the size-distribution of dust grains whose effect should be studied. 
We expect the density of dust layer becomes critical density for gravitational instability with the help of turbulence. 
However the dust grains in the turbulent flow have the velocity dispersion. Thus the critical density for gravitational instability possibly changes owing to the increased velocity dispersion of dust grains. We have to study the criterion of gravitational instability in more detail when the dust grains have the velocity dispersion in turbulent flows (c.f. Weidenschilling 1995). In subsequent papers, we plan to study these effects.

\acknowledgments{
The authors thank the anonymous referee for valuable comments and suggestion
that improved the paper. This work is supported by the Grant-in-Aid 
(No.15740118, 16077202, 16244120) 
from the Ministry of Education, Culture, Sports, Science, and 
Technology (MEXT) of Japan.}

\appendix
\clearpage
\section{The effect of the velocity dispersion of the dust grains}
We consider the velocity dispersion of the dust grains.
For the sake of simplicity, we assume that dust component obeys the equation of state of an ideal gas,
\begin{equation}
P_\mathrm{d} = c_\mathrm{d}^2 \rho_\mathrm{d},
\end{equation}
where $P_\mathrm{d}$ is the dust pressure, $c_\mathrm{d}$ the dust isothermal sound velocity. The equations governing this system are
\begin{equation}
 \nabla \cdot \mathbf v_{\mathrm{g}} =0,
\end{equation}
\begin{equation}
\dif{\rho_{\rm{d}}}{t} + \nabla (\rho_{\rm{d}} \mathbf v_{\mathrm{d}}) =0,
\end{equation}
\begin{equation}
\rho_{\rm{g}} \left(\dif{\mathbf v_{\mathrm{g}}}{t}+(\mathbf v_{\mathrm{g}} \cdot \nabla)\mathbf v_{\mathrm{g}} \right)
=- \nabla P_g+ A \rho_{\rm{g}} \rho_{\rm{d}} (\mathbf v_{\mathrm{d}} - \mathbf v_{\mathrm{g}}) ,
\end{equation}
\begin{equation}
\rho_{\rm{d}} \left(\dif{\mathbf v_{\mathrm{d}}}{t}+(\mathbf v_{\mathrm{d}} \cdot \nabla)\mathbf v_{\mathrm{d}} \right)
= -c_{\rm{d}}^2 \nabla \rho_d -A \rho_{\rm{g}} \rho_{\rm{d}} (\mathbf v_{\mathrm{d}} - \mathbf v_{\mathrm{g}}).
\end{equation}

We consider a Heviside step function velocity profile, and uniform density as the unperturbed state:
\begin{equation}
U(z)=
\left\{
	\begin{array}{rl}
		U_- & \mbox{for $z < 0 $} \\
		U_+ & \mbox{for $z > 0 $} 
	\end{array}
\right. .
\end{equation}
We restrict ourselves to the mode in the direction of the y-axis.
The hydrodynamic equations are given by
\begin{equation}
ikv_{\mathrm{g} \mathit{y}1}(z)+ \frac{dv_{\mathrm{g} \mathit{z}1}}{dz}(z)=0,
\end{equation}
\begin{equation}
- i (\omega- kU_{\pm} ) \rho_{\mathrm{d}1}(z) + i k \rho_{\mathrm{d}0} v_{\mathrm{d}\mathit{y}1}(z)  + \rho_{\mathrm{d}0} \frac{dv_{\mathrm{d}z1}}{dz}(z)=0,
\end{equation}
\begin{equation}
 -A \rho_{\mathrm{g}0} \rho_{\mathrm{d}0}(v_{\mathrm{d}x1}(z)-v_{\mathrm{g}x1}(z))-i\rho_{\mathrm{g}0} v_{\mathrm{g}x1}(z)( \omega - k U_{\pm})=0,
\end{equation}
\begin{equation}
-A \rho_{\mathrm{g}0} \rho_{\mathrm{d}0}(v_{\mathrm{d}\mathit{y}1}(z)-v_{\mathrm{g}y1}(z))-i\rho_{\mathrm{g}0} v_{\mathrm{g}y1}(z)( \omega - k U_{\pm}) +i k P_1(z)=0,
\end{equation}
\begin{equation}
-A \rho_{\mathrm{g}0} \rho_{\mathrm{d}0}(v_{\mathrm{d}\mathit{z}1}(z)-v_{\mathrm{g}z1}(z))-i\rho_{\mathrm{g}0} v_{\mathrm{g}z1}(z)( \omega - k U_{\pm}) + \frac{dP_1}{dz}(z)=0,
\end{equation}
\begin{equation}
-A \rho_{\mathrm{g}0} \rho_{\mathrm{d}0}(v_{\mathrm{g}x1}(z)-v_{\mathrm{d}x1}(z))-i\rho_{\mathrm{d}0} v_{\mathrm{d}x1}(z)( \omega - k U_{\pm})=0, 
\end{equation}
\begin{equation}
-A \rho_{\mathrm{g}0} \rho_{\mathrm{d}0}(v_{\mathrm{g}y1}(z)-v_{\mathrm{d}\mathit{y}1}(z))-i\rho_{\mathrm{d}0} v_{\mathrm{d}\mathit{y}1}(z)( \omega - k U_{\pm}) + i k c_{\mathrm{d}}^2 \rho_{\mathrm{d}1}(z)=0,
\end{equation}
\begin{equation}
-A \rho_{\mathrm{g}0} \rho_{\mathrm{d}0}(v_{\mathrm{g}\mathit{z}1}(z)-v_{\mathrm{d}\mathit{z}1}(z))-i\rho_{\mathrm{d}0} v_{\mathrm{d}\mathit{z}1}(z)( \omega - k U_{\pm}) + c_{\mathrm{d}}^2 \frac{d \rho_{\mathrm{d}1}}{dz}(z)=0.
\end{equation}
We transform the physical quantities to nondimensional ones by taking $k^{-1}$, $\rho_{\mathrm{g}0} k^{-3}$, and $(k U)^{-1}$ as units of the length, the mass and the time, respectively, where $U=U_+ - U_-$.

We rearrange the equations to obtain the matrix form of the linear equations:
\begin{equation}
\frac{d}{dz} \left(
  \begin{array}{c}
    P_1(z)      \\
    v_{\mathrm{g}\mathit{z}1}(z)  \\
   \rho_{\mathrm{d}1}(z) \\
   v_{\mathrm{d}\mathit{z}1}(z)   \\
  \end{array}
\right)= 
\left(
  \begin{array}{cccc}
     0  & -\frac{\mathcal{R}_\mathrm{d}}{t_{\mathrm{drag}}}+i (\omega-U_0) & 0 & \frac{\mathcal{R}_\mathrm{d}}{t_{\mathrm{drag}}}   \\
    \frac{1-i t_{\mathrm{drag}} (\omega-U_0)}{D}   &  0  &  \frac{c_{\mathrm{d}}^2}{D}  &  0  \\
     0  &  \frac{\mathcal{R}_\mathrm{d}}{c_{\mathrm{d}}^2 t_{\mathrm{drag}}}  & 0  & \frac{i \mathcal{R}_\mathrm{d} (i+t_{\mathrm{drag}}(\omega-U_0))}{c_{\mathrm{d}}^2 t_{\mathrm{drag}}}   \\
   \frac{1}{ D}    & 0   &  -\frac{N}{\mathcal{R}_\mathrm{d} D}  &  0  \\
  \end{array}
\right)
\left(  \begin{array}{c}
    P_1(z)      \\
    v_{\mathrm{g}\mathit{z}1}(z)  \\
   \rho_{\mathrm{d}1}(z) \\
   v_{\mathrm{d}\mathit{z}1}(z)   \\
  \end{array}
  \right)
  \label{eq:zdep},
\end{equation}
where
\begin{eqnarray}
N&=&-c_{\mathrm{d}}^2(\mathcal{R}_\mathrm{d}-i t_{\mathrm{drag}}(\omega-U_0))+(1+\mathcal{R}_\mathrm{d}-i t_{\mathrm{drag}}(\omega-U_0))(\omega-U_0)^2 \\
D&=&(i(1+\mathcal{R}_\mathrm{d})+t_{\mathrm{drag}}(\omega-U_0))(\omega-U_0)
\end{eqnarray}
We define the vector and the matrix as $\mathbf v$ and $\mathbf A$ respectively.

$\mathbf{v}$ is expressed as the linear combination of $\exp(\lambda_i z) \mathbf v_i$ where $i=1,2,3,4$. $\lambda_i$ and $\mathbf v_i$ are given by
\begin{eqnarray}
&&(\lambda_1, \lambda_2, \lambda_3, \lambda_4)=(1,-1, \lambda_\mathrm{a},-\lambda_\mathrm{a}), \\
&&(\mathbf v_1, \mathbf v_2, \mathbf v_3, \mathbf v_4)=\nonumber \\
&& \quad 
\left(
\left(  \begin{array}{c}
    1      \\
    \frac{(1- i t_{\mathrm{drag}} (\omega-U_0))}{D}  \\
    0 \\
    \frac{ 1 }{D}   \\
  \end{array}  \right),
\left(  \begin{array}{c}
    1      \\
    -\frac{(1- i t_{\mathrm{drag}} (\omega-U_0))}{D}  \\
    0 \\
    -\frac{ 1 }{D}   \\
  \end{array}  \right),
\left(  \begin{array}{c}
    1    \\
    0 \\
    -\frac{i  (\lambda_{\mathrm{a}}^2-1)t_{\mathrm{drag}}}{\omega-U_0} \\
    \frac{ \lambda_{\mathrm{a}} t_{\mathrm{drag}}}{\mathcal{R}_\mathrm{d}}   \\
  \end{array}  \right),
\left(  \begin{array}{c}
    1    \\
    0 \\
    -\frac{i  (\lambda_{\mathrm{a}}^2-1)t_{\mathrm{drag}}}{\omega-U_0} \\
    -\frac{ \lambda_{\mathrm{a}} t_{\mathrm{drag}}}{\mathcal{R}_\mathrm{d}}   \\
  \end{array}  \right)
\right) 
  , \nonumber \\
\end{eqnarray}
where
\begin{equation}
\lambda_\mathrm{a}=\frac{\sqrt{-t_{\mathrm{drag}}(-c_{\mathrm{d}}^2+(\omega-U_0)^2)-i(\omega-U_0)}}{c_{\mathrm{d}} \sqrt{t_{\mathrm{drag}}}}.
\end{equation}

We cannot determine the sign of $\mathrm{Re} (\lambda_\mathrm{a})$ if we do not know $\omega$. However if we assume $\mathrm{Re} (\lambda_\mathrm{a})>0$ to calculate $\omega$, we can obtain the consistent solutions. Thus we assume $\mathrm{Re} (\lambda_\mathrm{a})>0$. (In addition, in the case of $c_\mathrm{d}=0$ and $t_{\mathrm{drag}}=0$, we can show $\lambda_{\mathrm{a}}>0$.)

We seek the solutions which does not diverge in the limit of $|z| \to \infty$.
In the case where  $z>0$, the coefficients of $\mathbf v_1$ and $\mathbf v_3$ are zero.
In the case where $z<0$, the coefficients of $\mathbf v_2$ and $\mathbf v_4$ are zero.

Thus we have the conditions:
\begin{eqnarray}
P_{1+}+\frac{i(-\frac{1}{2}+\omega)(i(1+\mathcal{R}_\mathrm{d}) + (-\frac{1}{2}+\omega)t_{\mathrm{drag}})v_{\mathrm{g}z+}}{i+(-\frac{1}{2}+\omega)t_{\mathrm{drag}}}-\frac{i(-\frac{1}{2}+\omega)\rho_{\mathrm{d}+}}{t_{\mathrm{drag}}(-1+\lambda_{\mathrm{a}+}^2)}&=&0, \\
\frac{\mathcal{R}_\mathrm{d} v_{\mathrm{d}z+}}{\lambda_{\mathrm{a}+}}-\frac{i \mathcal{R}_\mathrm{d} v_{\mathrm{g}z+}}{(i+(-\frac{1}{2}+\omega)t_{\mathrm{drag}})\lambda_{\mathrm{a}+}}+\frac{i(-\frac{1}{2}+\omega)\rho_{\mathrm{d}+}}{-1+\lambda_{\mathrm{a}+}^2}&=&0, \\
P_{1-}+\frac{(\frac{1}{2}+\omega)(1+\mathcal{R}_\mathrm{d}-i (\frac{1}{2}+\omega)t_{\mathrm{drag}})v_{\mathrm{g}z-}}{i+(\frac{1}{2}+\omega)t_{\mathrm{drag}}}-\frac{i(\frac{1}{2}+\omega)\rho_{\mathrm{d}-}}{t_{\mathrm{drag}}(-1+\lambda_{\mathrm{a}-}^2)}&=&0, \\
-\frac{\mathcal{R}_\mathrm{d} v_{\mathrm{d}z-}}{\lambda_{\mathrm{a}-}}+\frac{2i \mathcal{R}_\mathrm{d} v_{\mathrm{g}z-}}{(2i+(1+2\omega)t_{\mathrm{drag}})\lambda_{\mathrm{a}-}}+\frac{i(\frac{1}{2}+\omega)\rho_{\mathrm{d}-}}{-1+\lambda_{\mathrm{a}-}^2}&=&0.
\end{eqnarray}
The boundary conditions are determined by the continuity of pressure and $v_{\mathrm{g}z1}/(\omega-k U)$, we have
\begin{equation}
P_{1+}=P_{1-},
\end{equation}
\begin{equation}
\rho_{\mathrm{d}1+}=\rho_{\mathrm{d}1-},
\end{equation}
\begin{equation}
\frac{v_{\mathrm{d}z-}}{\omega+\frac{1}{2}}=\frac{v_{\mathrm{d}z+}}{\omega-\frac{1}{2}},
\end{equation}
\begin{equation}
\frac{v_{\mathrm{g}z-}}{\omega+\frac{1}{2}}=\frac{v_{\mathrm{g}z+}}{\omega-\frac{1}{2}}.
\end{equation}

These equations have a nontrivial solution, if the determinant of the matrix of the linear equations is zero, thus we have the dispersion relation:
\begin{eqnarray}
&&(16 {t_{\mathrm{drag}}}^2 \omega^4+16 i (2+\mathcal{R}_\mathrm{d}){t_{\mathrm{drag}}} \omega^3-16(1+\mathcal{R}_\mathrm{d})\omega^2-4i(\mathcal{R}_\mathrm{d}-2){t_{\mathrm{drag}}} \omega -4\mathcal{R}_\mathrm{d}-4-{t_{\mathrm{drag}}}^2)
 \nonumber\\
&&\times
((2\omega-1)({t_{\mathrm{drag}}}(2\omega-1)+2i)\lambda_{\mathrm{a}-} - (2\omega+1)({t_{\mathrm{drag}}}(2\omega+1)+2i)\lambda_{\mathrm{a}+}
)
\nonumber \\
&&\qquad +
8\mathcal{R}_\mathrm{d} \left(  4{\omega }^2 -1\right) {c_{\mathrm{d}}}
   {{t_{\mathrm{drag}}}} =0.
\end{eqnarray}
In the limit of $c_{\mathrm{d}} \to 0$, this dispersion relation corresponds to the dispersion relation in Section \ref{sec:KH1}. Thus we can expect that the analysis in Section \ref{sec:KH1} is valid although $v_{\mathrm{d}z1}(z)/(\omega-kU)$ has the discontinuity at the boundary.  $v_{\mathrm{d}z1}(z)/(\omega-kU)$ becomes discontinuous at the boundary in the case where $c_{\mathrm{d}}=0$ although that is continuous for any case where $c_{\mathrm{d}}\ne0$.

\begin{figure}[h]
\plotone{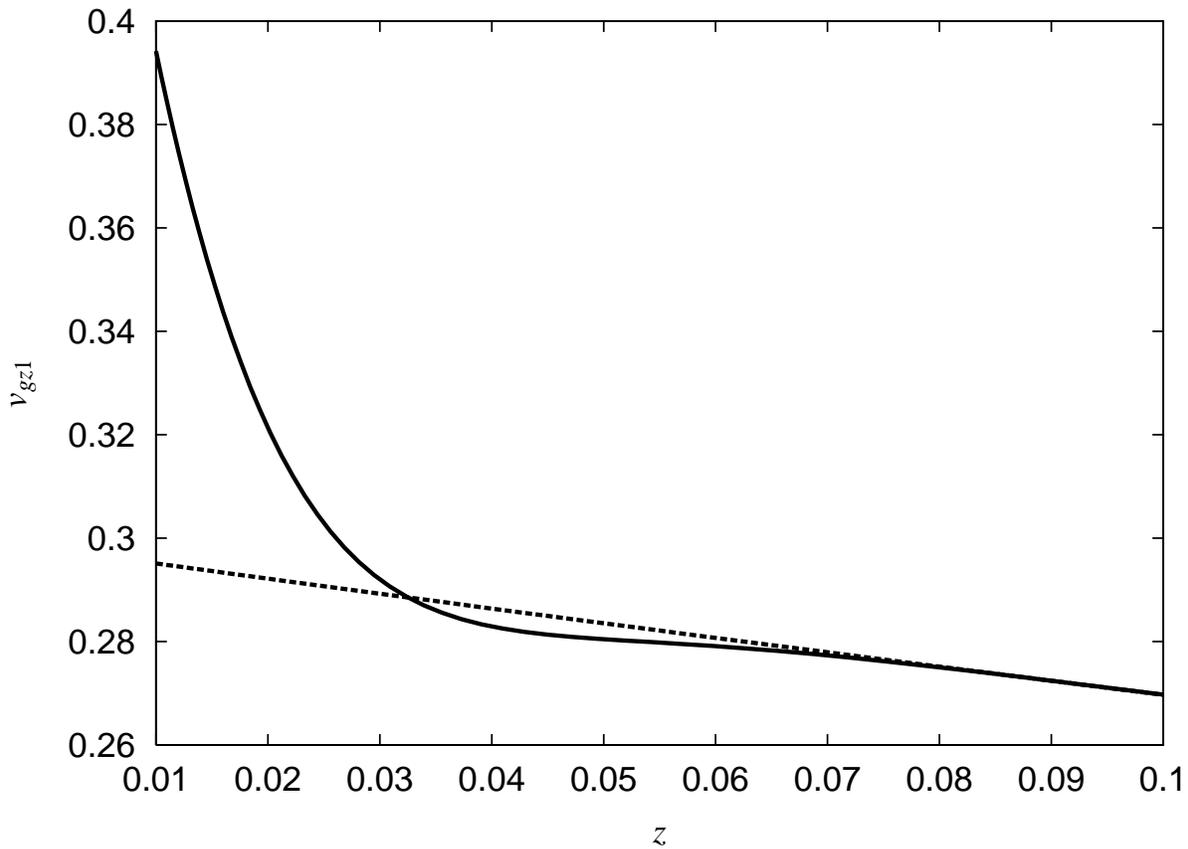}
\caption{The amplitude of the eigenfunction $v_{\mathrm{d}\mathit{z}1}(z)$ in the case where $c_{\mathrm{d}} = 1$ (solid line) and the amplitude of the eigenfunction $v_{\mathrm{d}\mathit{z}1}(z)=C_1 \exp(-z)$ in the case where $c_{\mathrm{d}} = 0$ (dotted line).}
\label{fig:vgz.eps}
\end{figure}

\begin{figure}
\plotone{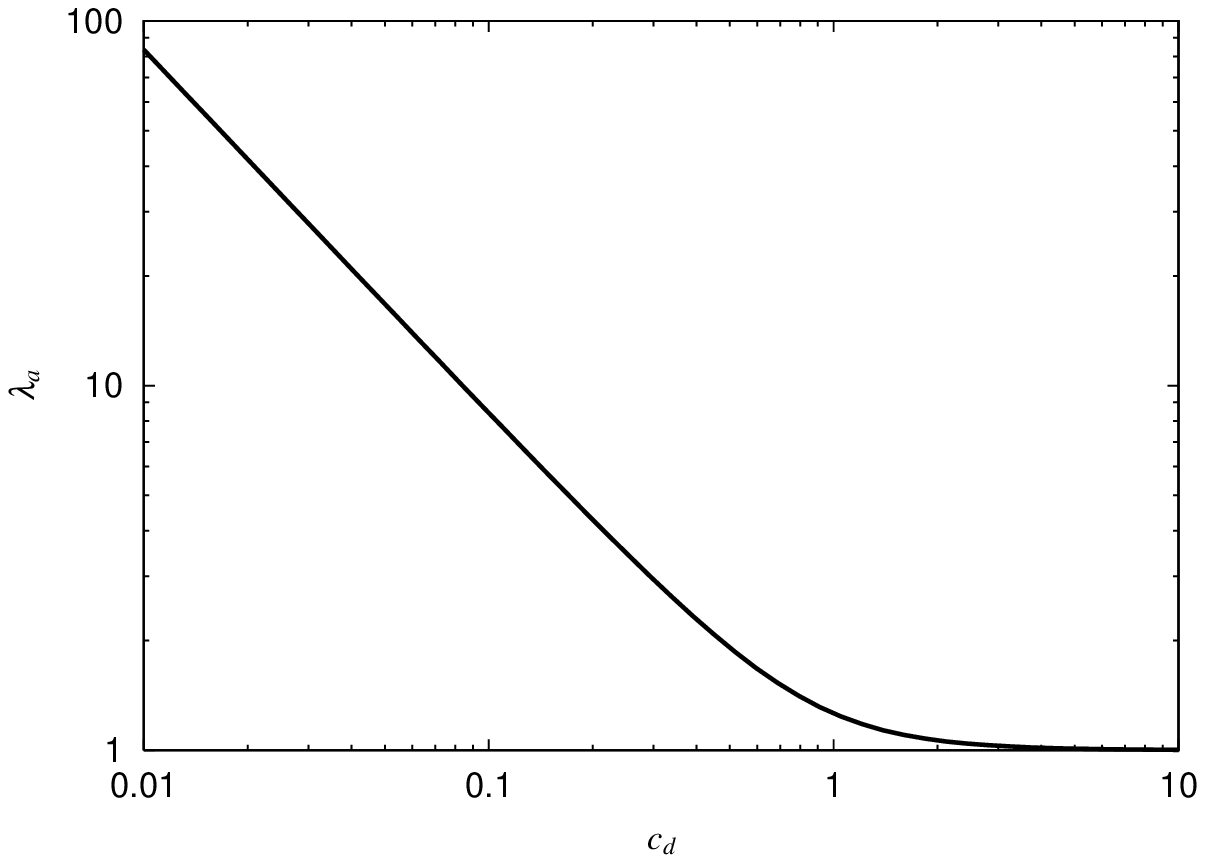}
\caption{The real part of $\lambda_{\mathrm{a}}$ as a fuanction $c_{\mathrm{d}}$ in the case where $t_{\mathrm{drag}}=1,\mathcal{R}_\mathrm{d}=1$.}
\label{fig:lambda.eps}
\end{figure}

To understand the reason of this transition from continuity to discontinuity in the limit of $c_{\mathrm{d}} \to 0$, we investigate the velocity eigenfunction of dust.
The velocity perturbation of dust is given by 
\begin{equation}
v_{\mathrm{d}\mathit{z}1}(z)=C_1 \exp(-z) + C_2 \exp(-\lambda_{\mathrm{a}} z),
\end{equation}
where $C_1$ and $C_2$ are constant. 
This is shown in Figure \ref{fig:vgz.eps}.
The dotted line corresponds to the case where $c_{\mathrm{d}}\to0$.
The correction to the $c_{\mathrm{d}}\to0$ case is $C_2 \exp(-\lambda_{\mathrm{a}} z)$.
$v_{\mathrm{d}z1}(z)/(\omega-kU)$ is continuous because of this correction, that is,
if we consider only $C_1 \exp(-z)$, $v_{\mathrm{d}z1}(z)/(\omega-kU)$ is discontinuous at the boundary.
This correction term becomes significant in the region where $|z| \lesssim \lambda$.
Figure \ref{fig:lambda.eps} shows that $\lambda$ diverges in the limit of $c_{\mathrm{d}}\to0$, thus the width of the region for the correction decreases to $0$.
Hence the correction term disappears effectively. In this way we can understand the transition from continuity to discontinuity in the limit of $c_{\mathrm{d}} \to 0$.

\begin{figure}
\plotone{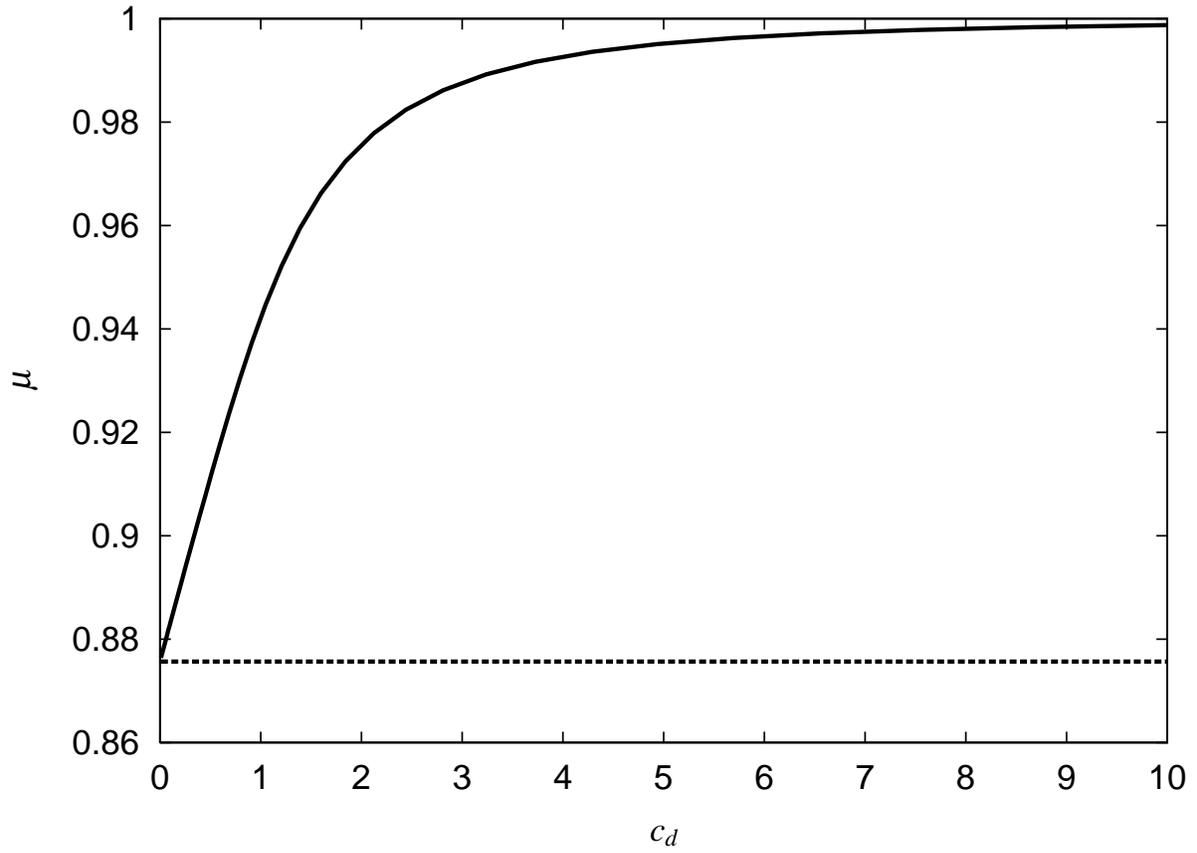}
\caption{The the growth rate as a function of $c_{\mathrm{d}}$.
The solid line shows the growth rate where $c_{\mathrm{d}}=1$ and the dotted line shows the growth rate where $c_{\mathrm{d}}=0$.}
\label{fig:mu_A.eps}
\end{figure}

In contrast, Figure \ref{fig:mu_A.eps} shows that the growth rate converges to that of $c_{\mathrm{d}}=0$ in the limit of $c_{\mathrm{d}} \to 0$.  
Thus our result from the analysis with discontinuous velocity perturbation for dust with $c_{\mathrm{d}}=0$ can be regarded as a rigorous result for the case where $c_{\mathrm{d}}$ is very small.

\begin{figure}
\plotone{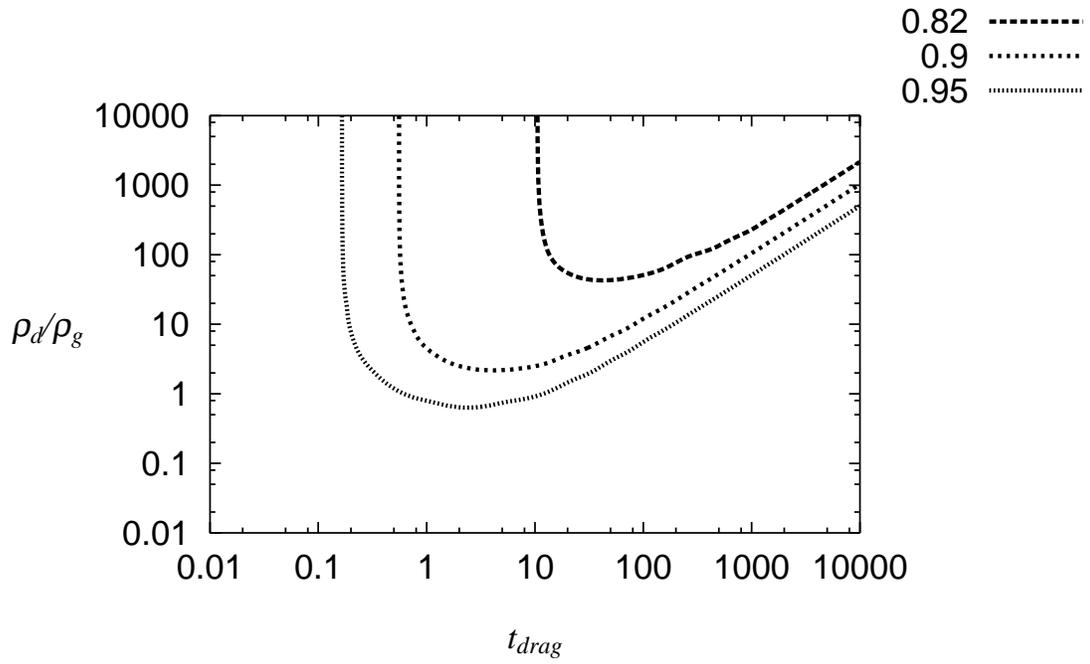}
\caption{The contour of growth rate as a function of $t_{\mathrm{drag}}$ and $\mathcal{R}_\mathrm{d}$ in the case where $c_{\mathrm{d}}=1$.}
\label{fig:mucdcontour_A.eps}
\end{figure}

\begin{figure}
\plotone{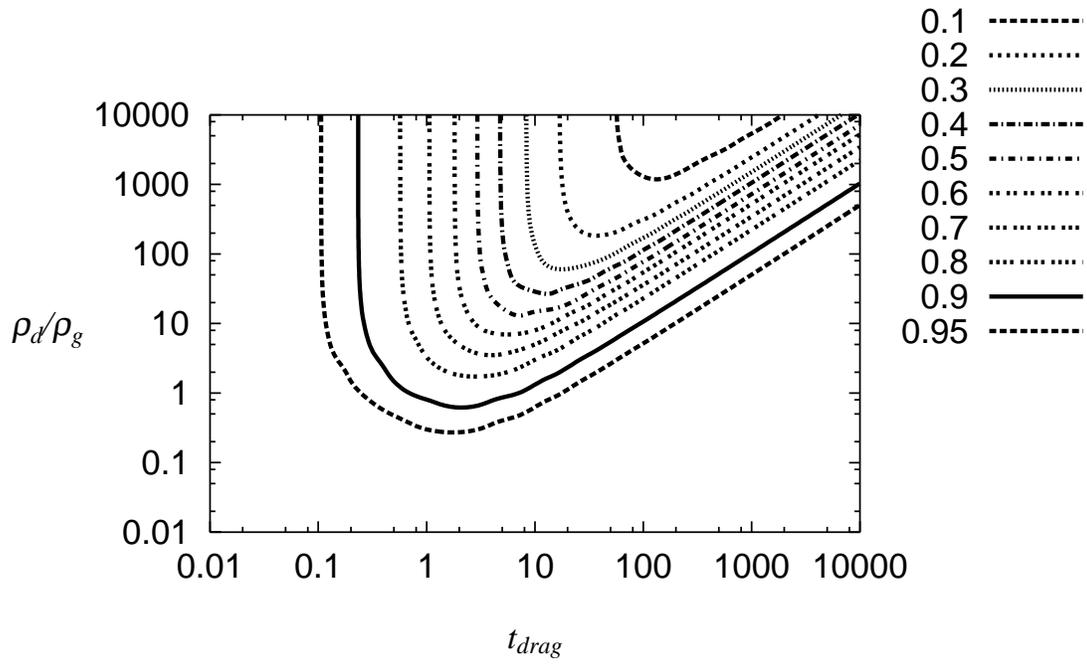}
\caption{Same as Fig.\ref{fig:mucdcontour_A.eps} but for $c_{\mathrm{d}}=0.1$.}
\label{fig:mucdcontour_B.eps}
\end{figure}

In the limit of $c_{\mathrm{d}}\to \infty$,
the growth rate approaches to that of simple Kelvin-Helmholtz instability.
As shown in Figure \ref{fig:mucdcontour_A.eps} and Figure \ref{fig:mucdcontour_B.eps},
the stabilization by the friction lessens as $c_{\mathrm{d}}$ increases. 


\section{The most unstable mode \label{apd:appr}}
We show that the solution of the equation (\ref{eq:dr1}) which has the largest real part is real.

We define the left hand side of the equation (\ref{eq:dr1}) as $f(\tilde \omega)$.
We can show $f(0)<0$, $f( \infty) = \infty$, $f( -\infty) = \infty$.
It follows that the equation (\ref{eq:dr1}) must have at least one positive solution $x_1$ and one negative solution $x_2$.
If two other solutions are real, it is true that the solution that has the largest real part is real.
In following discussion, we consider the case that the equation (\ref{eq:dr1}) has two complex solutions.

The complex solutions are $x_3$ and its complex conjugate $x_3^*$ because all coefficients are real numbers.
We can rewrite the equation:
\begin{equation}
(\omega^2 -2 l \omega -m^2)(\omega^2-2\gamma \omega+n^2)=0,
\end{equation}
where
\begin{equation}
m=\sqrt{-x_1 x_2},
\end{equation}
\begin{equation}
n=|x_3|,
\end{equation}
\begin{equation}
l=x_1+x_2,
\end{equation}
\begin{equation}
\gamma = \mathrm{Re} (x_3),
\end{equation}
This equation is equivalent to the equation (\ref{eq:dr1}), we obtain the equations:
\begin{equation}
-(l+\gamma)=\frac{2+\mathcal{R}_\mathrm{d} }{t_{\mathrm{drag}}},
\label{eq:3dod1}
\end{equation}
\begin{equation}
-m^2+n^2+4 l \gamma= \frac{4+4\mathcal{R}_\mathrm{d}}{t_{\mathrm{drag}}^2},
\label{eq:3dod2}
\end{equation}
\begin{equation}
-l n^2+m^2 \gamma= \frac{-2 t + \mathcal{R}_\mathrm{d} t_{\mathrm{drag}}}{t_{\mathrm{drag}}},
\label{eq:3dod3}
\end{equation}
\begin{equation}
m^2 n^2 = \frac{4+4\mathcal{R}_\mathrm{d}+t_{\mathrm{drag}}^2}{t_{\mathrm{drag}}^2}.
\label{eq:3dod4}
\end{equation}

From the equation (\ref{eq:3dod1}) and (\ref{eq:3dod3}), we give
\begin{equation}
l=-\frac{-2+\mathcal{R}_\mathrm{d}+m^2(2+\mathcal{R}_\mathrm{d})}{(m^2+n^2)t_{\mathrm{drag}}},
\label{eq:apBc}
\end{equation}
\begin{equation}
\gamma=\frac{-2+\mathcal{R}_\mathrm{d}-n^2(2+\mathcal{R}_\mathrm{d})}{(m^2+n^2)t_{\mathrm{drag}}}.
\label{eq:apBa}
\end{equation}

In the case of $\mathcal{R}_\mathrm{d} \le 2$, from the equation (\ref{eq:apBa}), we can see $\gamma<0$ .

In the case of $\mathcal{R}_\mathrm{d}>2$, we give the inequality $l<0$ from the equation (\ref{eq:apBc}).
Here after we use the reduction to absurdity to prove $\gamma < 0$.
We suppose $\gamma>0$. From the equation (\ref{eq:3dod2}), it can be seen that $-m^2+n^2>0$.
From the equation (\ref{eq:3dod4}) we can see $m^2n^2>1$. Then we can see $n^2>1$.
On the other hand, from the equation (\ref{eq:apBa}) we give the inequality
\begin{equation}
\frac{\mathcal{R}_\mathrm{d}-2}{\mathcal{R}_\mathrm{d}+2}>n^2.
\end{equation}
It can be seen that $\displaystyle{\frac{\mathcal{R}_\mathrm{d}-2}{\mathcal{R}_\mathrm{d}+2}}<1$.
This implies that $n^2<1$, a contradiction. Thus we obtain the inequality $\gamma<0$.

This means that the real parts of the complex solutions are always negative. Thus the solution that has the largest real part is not complex but real: the most unstable mode is the exponentially growing solution.


\begin{thebibliography}{}
\bibitem[Adachi, Hayashi, \& Nakazawa(1976)]{Adachi1976} Adachi, 
I., Hayashi, C., \& Nakazawa, K.\ 1976, Progress of Theoretical Physics, 
56, 1756 
\bibitem[Champney et al.(1995)]{Champney1995}Champney, J. M., A. R. Dobrovolskis, and J. N. Cuzzi, 1995, Phys. Fluids, 7, 1703.
\bibitem[Chandrasekhar(1961)]{Chandrasekhar1961}Chandrasekhar, S. 1961, Hydrodynamic and Hydromagnetic Stability (Oxford:Oxford University Press).
\bibitem[Coradini, Magni, \& Federico(1981)]{Coradini1981} Coradini, 
A., Magni, G., \& Federico, C.\ 1981, \aap, 98, 173 
\bibitem[Cuzzi et al.(1993)]{Cuzzi1993} Cuzzi, J. N., Dobrovolskis, A. R., \& Champney, J. M. 1993, Icarus, 106, 102.
\bibitem[Dobrovolskis et al.(1999)]{Dobrovolskis1999}Dobrovolskis, A. R., J. S. Dacles-Mariani, and J. N. Cuzzi, 1999, J. Geophys. Res., 104, 30805.
\bibitem[Garaud \& Lin(2004)]{Garaud2004} Garaud, P., \& Lin, D. N. C. 2004, \apj, 603, 292.
\bibitem[Goldreich \& Ward(1973)]{Goldreich1973} Goldreich, P.~\& Ward, W.~R.\ 1973, \apj, 183, 1051 
\bibitem[Hayashi(1981)]{Hayashi1981} Hayashi, C. 1981, Progr. Theor. Phys. Suppl. 70, 35.
\bibitem[Ishitsu \& Sekiya(2002)]{Ishitsu2002} Ishitsu, N., \& Sekiya, M. 2002, Earth Planets Space, 54, 917.
\bibitem[Ishitsu \& Sekiya(2003)]{Ishitsu2003} Ishitsu, N., \& Sekiya, M. 2003, Icarus, 165, 181.
\bibitem[Nakagawa et al.(1986)]{Nakagawa1986} Nakagawa, Y., Sekiya, M., \& Hayashi, C., 1986, Icarus, 67, 375,
\bibitem[Poppe et al.(2000)]{Poppe2000} Poppe, T., Blum, J., \& Henning, T.\ 2000, \apj, 533, 454 
\bibitem[Safronov(1969)]{Safronov1969} Safronov, V. S. 1969, Evolution of the Protoplanetary Cloud and Formation of the Earth and the PLanets(Moscow: Nauka)
\bibitem[Sekiya(1983)]{Sekiya1983} Sekiya, M.\ 1983, Progress of Theoretical Physics, 69, 1116 
\bibitem[Sekiya(1998)]{Sekiya1998} Sekiya, M. 1998, Icarus, 133, 298.
\bibitem[Sekiya \& Ishitsu(2000)]{Sekiya2000} Sekiya, M., \& Ishitsu, N. 2000, Earth Planets Space, 52, 517.
\bibitem[Sekiya \& Ishitsu(2001)]{Sekiya2001} Sekiya, M., \& Ishitsu, N. 2001, Earth Planets Space, 53, 761.
\bibitem[Sekiya \& Takeda(2003)]{Sekiya2003} Sekiya, M., \& Takeda, H.\ 2003, Earth, Planets, and Space, 55, 263
\bibitem[Yamoto \& Sekiya(2004)]{Yamoto2004} Yamoto, F., \& Sekiya, M.\ 2004, Icarus, 170, 180 
\bibitem[Youdin \& Chiang(2004)]{Youdin2004} Youdin, A.~N.~\& Chiang, E.~I.\ 2004, \apj, 601, 1109 
\bibitem[Youdin \& Shu(2002)]{Youdin2002} Youdin, A. N., \& Shu, F. H. 2002, \apj, 580, 494.
\bibitem[Weidenschilling(1977)]{Weidenschilling1977} Weidenschilling, S.~J.\ 1977, \mnras, 180, 57 
\bibitem[Weidenschilling(1980)]{Weidenschilling1980} Weidenschilling, S. J., 1980, Icarus, 44, 172.
\bibitem[Weidenschilling(1995)]{Weidenschilling1995} Weidenschilling, S.~J.\ 1995, Icarus, 116, 433 


\end{thebibliography}
\end{document}